%% file: arxiv2.tex
\documentclass{old-jaa}
\usepackage[british]{babel}             
\usepackage{newtxtext}                  
\usepackage[slantedGreek]{newtxmath}    
\usepackage[T1]{fontenc}                
\usepackage{graphicx}                   
\usepackage[authoryear,longnamesfirst]{natbib}
\bibpunct{(}{)}{;}{a}{}{,}
\usepackage{hyperref}
\hypersetup{pdfauthor={D. A. Green},
            pdftitle={A revised catalogue of 294 Galactic supernova remnants},
            pdfkeywords={supernova remnants, catalogues, ISM: general},
            bookmarksnumbered=true}
\usepackage{fix-jaa}
\def\SNR(#1)#2(#3){{G#1$#2$#3}}
\newcommand{\SigmaUnit}{{\rm W~m^{-2}\,Hz^{-1}\,sr^{-1}}}
\def\fdeg{^\circ\mkern-7mu.\mkern1mu}
\begin{document}\sloppy
\doinum{}
\artcitid{(in press)}
\volnum{40}
\year{2019}
\setcounter{page}{1}

\title{A revised catalogue of 294 Galactic supernova remnants}
\author{D.~A.\ Green}
\affilOne{\rm Astrophysics Group, Cavendish Laboratory,
       19 J.~J.~Thomson Avenue, Cambridge CB3 0HE, United Kingdom}

\twocolumn[{

\maketitle

\vskip 0pt\noindent\hspace{15mm}\begin{minipage}{16cm}{\fontsize{10}{12.5}\selectfont
E-mail: D.A.Green@mrao.cam.ac.uk}\end{minipage}\vskip 15pt

\msinfo{11 June 2019}{}

\label{firstpage}

\begin{abstract}
A revised catalogue of Galactic supernova remnants (SNRs) is presented,
along with some simple statistics of their properties. Six new SNRs have
been added to the catalogue since the previous published version from 2014,
and six entries have been removed, as they have been identified as
{\sc H\,ii} regions, leaving the number of entries in the catalogue at
294. Some simple statistics of the remnants in the catalogue, and the
selection effects that apply, are discussed, along with some recently
proposed Galactic SNR candidates.
\end{abstract}

\keywords{supernova remnants --- catalogues --- ISM: general}

}]

\def\snrN{294}
\section{Introduction}\label{s:intro}

This paper presents the latest version of  a catalogue of Galactic
supernova remnants (SNRs) which I have compiled for several decades.
Previous versions have been published in \citep{1984MNRAS.209..449G,
1988Ap&SS.148....3G, 1991PASP..103..209G, 2002ISAA....5.....S,
2004BASI...32..335G, 2009BASI...37...45G, 2014BASI...42...47G}. In
addition, more detailed web-based versions of the catalogue have been
produced since 1995 -- most recently in 2017 -- which either correspond
to one of the published catalogues, or are an intermediate revision.
This version of the catalogue contains {\snrN} entries.
Section~\ref{s:catalogue} gives the details of the entries in the
catalogue, and Section~\ref{s:new} discusses the entries added or
removed from the catalogue since the last published version
\citep{2014BASI...42...47G}. Section~\ref{s:discuss} discusses some
simple statistics of the remnants in the current catalogue, the
selection effects that apply to the identification of Galactic SNRs, and
some recently proposed candidate SNRs.

\section{The catalogue format}\label{s:catalogue}

This catalogue is based on the literature published up to the end of
2018, and contains {\snrN} entries. For each SNR in the catalogue the
following parameters are given.
\begin{itemize}
\item {\bf Galactic Coordinates} of the remnant. These are quoted to a
tenth of a degree, as is conventional. In this catalogue additional
leading zeros are not used. These are generally taken from the Galactic
coordinate based name used for the remnant in the literature. It should
be noted that when these names were first defined, they may not follow
the IAU recommendation\altfoonoterule\footnote{see: {\tt
http://cdsweb.u-strasbg.fr/Dic/iau-spec.htx}} that coordinates should be
truncated, not rounded to construct such names.
\item {\bf Right Ascension} and {\bf Declination} of J2000.0 equatorial
coordinates of the source centroid, for which an accuracy of the quoted
values depends on the size of the remnant. For small remnants they are
to the nearest few seconds of time and the nearest minute of arc
respectively, whereas for larger remnants they are rounded to coarser
values, but are in every case sufficient to specify a point within the
boundary of the remnant. These coordinates are usually deduced from
radio images rather than from X-ray or optical observations.
\item {\bf Angular Size} of the remnant, in arcminutes. This is usually
taken from the highest resolution radio image available. The boundary of
most remnants approximates reasonably well to either a circle or to an
ellipse. A single value is quoted for the angular size of the more
nearly circular remnants, which is the diameter of a circle with an area
equal to that of the remnant. For more elongated remnants the product of
two values is given, which are the major and minor diameters of the
remnant boundary modelled as an ellipse. In a small number of cases an
ellipse is not a good description of the boundary of the object (which
will be noted in the description of the object given in its catalogue
entry), although an angular size is still quoted for information. For
`filled-centre' type remnants (see below), the size quoted is for the
largest extent of the observed emission, not, as at times has been used
by others, the half-width of the centrally brightened peak.
\item {\bf Type} of the SNR: `S' or `F' if the remnant shows a `shell'
or `filled-centre' structure, or `C' if it shows `composite' (or
`combination') radio structure, with a combination of shell and
filled-centre characteristics. If there is some uncertainty, the type is
given as `S?', `F?' or `C?', and as `?' in several cases where an object
is conventionally regarded as an SNR even though its nature is poorly
known or it is not well-understood. (Note: the term `composite' has been
used, by some authors, in a different sense, to describe remnants with
radio shell and centrally-brightened X-ray emission. An alternative term
used to describe such remnants is `mixed morphology', e.g.\ see
\citealt{1998ApJ...503L.167R}.)
\item {\bf Flux Density} of the remnant at a frequency of 1~GHz, in
jansky. This is {\sl not} a measured value, but is instead derived from
the observed radio spectrum of the source. The frequency of 1~GHz is
chosen because flux density measurements are usually available at both
higher and lower frequencies. Some young remnants -- notably
\SNR(111.7)-(2.1) ($=$Cassiopeia A) and \SNR(184.6)-(5.8) ($=$Crab
Nebula), but also \SNR(130.7)+(3.1) ($=$3C58) and \SNR(120.1)+(1.4)
($=$Tycho) -- show secular variations in their radio flux density. In
this revision of the catalogue the 1-GHz flux densities for
\SNR(111.7)-(2.1) and \SNR(184.6)-(5.8) have been taken from
\citep{2017ApJS..230....7P}, for an epoch of 2016. Results from the
primary literature should be used for any detailed quantitative studies
of the radio spectra of these and other remnants.
\item {\bf Spectral Index} of the integrated radio emission from the
remnant, $\alpha$ (here defined in the sense, $S \propto \nu^{-\alpha}$,
where $S$ is the flux density at frequency $\nu$). This is either a
value that is quoted in the literature, or one deduced from the
available integrated flux densities of the remnant. For several SNRs a
simple power law is not adequate to describe their radio spectra, either
because there is evidence that the integrated spectrum is curved or the
spectral index varies across the face of the remnant. In these cases the
spectral index is given as `varies' (refer to the description of the
remnant and appropriate references in the detailed catalogue entry for
more information). In some cases, for example where the remnant is
highly confused with thermal emission, the spectral index is given as
`?' since no value can be deduced with any confidence. These spectral
indices have a very wide range of quality, and the primary literature
should be consulted for any detailed study of the radio spectral indices
of these remnants.
\item {\bf Other Names} that are commonly used for the remnant. Note
that these are given in parentheses if the remnant is only a part of the
source. For some well known remnants -- e.g.\ \SNR(184.6)-(5.8), the Crab
Nebula -- not all common names are given.
\end{itemize}
A summary of the data available for all {\snrN} remnants in the
catalogue is given in Table~\ref{t:snrcat}.

A more detailed version of the catalogue is available at:
 \\[6pt]
 \centerline{\tt http://www.mrao.cam.ac.uk/surveys/snrs/}
 \\[6pt]
In addition to the basic parameters which are given in Table~1, the
detailed catalogue contains the following additional information.
(i) Notes on the remnant. For example, if other Galactic coordinates
have at times been used to label it (usually before good observations
have revealed the full extent of the object, but sometimes in error); if
the SNR is thought to be the remnant of a historical SN.
(ii) Short descriptions of the observed structure/properties of the
remnant at radio, optical and X-ray wavelengths, as appropriate from
available observations.
(iii) Comments on distance determinations, and any point sources or
pulsars in or near the object (although they may not necessarily be
related to the remnant).
(iv) References to observations are given for each remnant, complete
with journal, volume, page, and a short description of what information
each paper contains (e.g.\ for radio observations these generally
include the telescopes used, the observing frequencies and resolutions,
together with any flux density determinations). These references are
{\sl not} complete, but cover recent and representative observations of
the remnant that are available, and should themselves include references
to earlier work. These references are from the published literature up
to the end of 2018.

The detailed version of the catalogue is available in pdf format for
downloading and printing, or as web pages, including a page for each
individual remnant. The web pages for each remnant include links to the
`NASA Astrophysics Data System' for each of the over three thousand
references that are included in the detailed listings for individual
SNRs.

Some of the parameters included in the catalogue are themselves of
variable quality. For example, the radio flux density of each remnant at
1~GHz is generally obtained from several radio observations over a range
of frequencies, both above and below 1~GHz, so is of good quality.
However, there are 21 remnants -- often those which have been identified
at other than radio wavelengths -- for which no reliable radio flux
density is yet available, because they have either not been detected or
well observed in the radio. Although the detailed version of the
catalogue contains notes on distances for many remnants reported in the
literature, these are highly variable in terms of reliability and
accuracy. Consequently the distances given within the detailed catalogue
should be used with caution in any statistical studies, and reference
should be made to the primary literature cited in the detailed
catalogue.

The detailed version of the catalogue also contains notes both on those
objects no longer thought to be SNRs, and on the many possible and
probable remnants that have been reported in the literature (including
possible large, old remnants, seen from radio continuum, X-ray or {\sc
H\,i} observations). See Section \ref{s:pp} below, for discussion of
some recently proposed remnants.

It should be noted that the catalogue is far from homogeneous. Although
many remnants, or possible remnants, were first identified from
wide-area radio surveys, there are many others that have been observed
with diverse observational parameters, making uniform criteria for
inclusion in the main catalogue difficult. For an alternative,
high-energy catalogue of SNRs see \citet{2012AdSpR..49.1313F}.

\section{SNRs added to/objects removed from the catalogue}\label{s:new}

Since the last published version \citep{2014BASI...42...47G}, the
following supernova remnants have been added to the catalogue.

\begin{itemize}
\item \SNR(351.0)-(5.4), which was identified by
\citep{2014A&A...568A.107D} from radio and other observations.
\item A very high-latitude remnant, \SNR(70.0)-(21.5), identified
primarily from optical observations by \citet{2015ApJ...812...37F}.
Previously \citet{2002A&A...396..225B} had noted optical filaments in
this region, which they suggested were indicative of one or more SNRs.
As noted by both \citeauthor{2002A&A...396..225B} and
\citeauthor{2015ApJ...812...37F} there is also faint X-ray emission from
this remnant.
\item \SNR(181.1)+(9.5), another high-latitude remnant, identified from
radio observations by \citet{2017A&A...597A.116K}.
\item \SNR(323.7)-(1.0), which was one of several candidate remnant
given by \citet{2014PASA...31...42G}, which was confirmed as a SNR from
$\gamma$-ray observations, see \citet{2017ApJ...843...12A} and
\citet{2018A&A...612A...8H}.
\item Possible faint radio SNRs near $l=150\fdeg5$, $b=4\fdeg0$ have
been reported by \citet{2014A&A...566A..76G} and
\citet{2014A&A...567A..59G}. \citeauthor{2014A&A...567A..59G} proposed a
large ($180 \times 150$ arcmin$^2$) remnant, \SNR(150.3)+(4.5), whereas
\citeauthor{2014A&A...566A..76G} proposed part of this as a smaller ($61
\times 18$ arcmin$^2$) remnant, \SNR(150.8)+(3.8). Recently
\citet{2018ApJS..237...32A} have shown extended $\gamma$-ray emission
from much of \SNR(150.3)+(4.5), confirming it as a SNR.
\item \SNR(53.4)+(0.0) was confirmed as a SNR by
\citet{2018ApJ...860..133D}, from radio and X-ray observations. This is
one of several candidate SNRs in this region (e.g.\
\citealt{2017A&A...605A..58A}). See also \citet{ 2018ApJ...866...61D}.
\end{itemize}
In this version of the catalogue five objects previously listed as SNRs
have been removed, namely (\SNR(20.4)+(0.1), \SNR(21.5)-(0.1),
\SNR(23.6)+(0.3), \SNR(59.8)+(1.2) and \SNR(65.8)-(0.5), as they have
been identified as {\sc H\,ii}  regions by \citet{2017A&A...605A..58A}.
Also, \SNR(192.8)-(1.1) has been removed, as \citet{2011A&A...529A.159G}
had shown that this is not a SNR (see also
\citealt{2014JKAS...47..259K}). Erroneously it was not removed from the
2014 version of the catalogue. Note that \SNR(358.1)+(1.0) was
erroneously labelled \SNR(358.1)+(0.1) in 2009 and 2014 versions of the
catalogue, which has now been corrected.

\section{Discussion}\label{s:discuss}

\subsection{Some Simple Statistics}


There are 21 Galactic SNRs which do not have a flux density at 1~GHz in
the catalogue. This is because either the remnant has not been detected
at radio wavelengths, or it is poorly defined by current radio
observations, so that their flux density at 1~GHz cannot be determined
with any confidence: i.e.\ 93\% of the remnants do have a flux density
at 1~GHz in the catalogue. Of the catalogued remnants, $\approx 42$\%
are detected in X-rays, and $\approx 31$\% in the optical. The smaller
proportion of SNR identified in the optical and X-ray wavebands is due
to Galactic absorption, which hampers the detection of distant remnants.

In this version of the catalogue, 80\% of remnants are classified as
shell (or possible shell) remnants, 13\% are composite (or possible
composite) remnants, and just 3\% are filled-centre (or possible filled
centre) remnants. The types of the remaining remnants are not clear from
current observations (or else they are objects which are conventionally
regarded as SNRs although they do not fit well into any of the
conventional types, e.g.\ CTB80 ($=$\SNR(69.0)+(2.7)), MSH 17$-$3{\em 9}
($=$\SNR(357.7)-(0.1))).

\subsection{Selection Effects}

In previous papers (e.g.\ \citealt{1991PASP..103..209G,
2005MmSAI..76..534G}) I have discussed the selection effects that apply
to the identification of Galactic SNRs. Although some SNR are identified
first at other than radio wavelengths, most SNRs have been identified
first in the radio. The selection effects for the SNR catalogue are
therefore dominated by those that apply at radio wavelengths. These are:
(i) the difficulty in finding low surface brightness remnants, and (ii)
the difficulty in finding small angular size remnants.

In \citet{2005MmSAI..76..534G} I derived a surface brightness
completeness limit of $\Sigma \approx 10^{-20}$ ${\SigmaUnit}$, at
1~GHz. This nominal completeness limit is supported by various searches
for SNRs, as no remnants with a surface brightness above this limit have
been added to the published versions of the catalogue since
\citet{2009BASI...37...45G}. \citet{2013A&A...559A..81X} used
multi-frequency radio observations to separate thermal and non-thermal
radio emission in a large region around Cygnus X ($66^\circ \le l \le
90^\circ$, $|b| < 4^\circ$). They did not find any new large SNRs with
$\Sigma > 0.37 \times 10^{-20}$ ${\SigmaUnit}$, consistent with
previously quoted the completeness limit. More recently
\citet{2017A&A...605A..58A} have identified many candidate SNRs in the
region $17\fdeg5 < l < 67\fdeg4$, $|b| \le 1\fdeg25$, from THOR (e.g.\
\citealt{2016A&A...595A..32B}) and VGPS (e.g.\
\citealt{2006AJ....132.1158S}) radio continuum observations at 1.4~GHz
and mid-IR surveys. This covers a large fraction of the inner Galaxy --
where most Galactic SNRs are expected to be -- but only 2 of these
candidates appear to have a surface brightness at 1~GHz above $10^{-20}$
${\SigmaUnit}$ (assuming a spectral index of 0.5 to scale the observed
1.4~GHz flux density to 1~GHz).

This surface brightness completeness limit of $10^{-20}$ ${\SigmaUnit}$
was used in \citet{2015MNRAS.454.1517G} to select a sample of 69 SNRs
from the 2014 version of the catalogue. This sample was then used to
derive constraints on the distribution of remnants with Galactocentric
radius. Of the six new remnants added to the catalogue since the 2014
version, only one (\SNR(53.4)+(0.0)) has an integrated flux density at
1~GHz, and it is fainter than $\Sigma \approx 10^{-20}$
${\SigmaUnit}$. The other five, are either not detected in the radio
(\SNR(70.0)-(21.5)) or do not currently have integrated radio flux
densities as they are faint. Of the objects removed from the catalogue
since 2014, two had a surface brightness of $1.2 \times 10^{-20}$ and
$2.1 \times 10^{-20}$ ${\SigmaUnit}$ (for \SNR(20.4)+(0.1) and
\SNR(23.6)+(0.3) respectively). So, the current catalogue contains 67
remnants with a surface brightness above $10^{-20}$ ${\SigmaUnit}$. This
is two less than in the sample used in \citet{2015MNRAS.454.1517G}, so
that the results derived there will not be significantly changed.

Small angular size remnants -- which will be the young but distant SNRs
in the Galaxy -- need to be resolved, for their structure to be
recognised. Most wide field radio surveys have not had small enough
resolutions to easily identify such small angular size remnants. There
are only 9 SNRs with angular diameters $\le 3$~arcmin in the catalogue,
and none of these have been added to the catalogue since the 2009
version of the catalogue.

\begin{figure}
\centerline{\includegraphics[width=8.5cm]{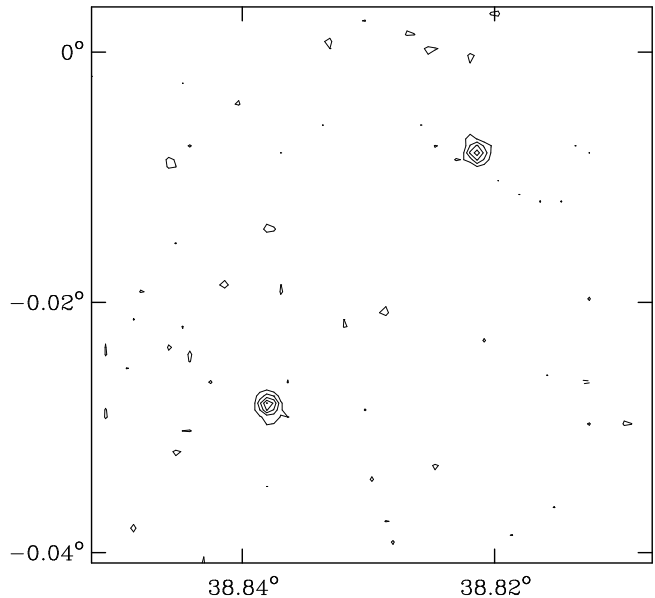}}
\caption{MAGPIS \citep{2006AJ....131.2525H} 1.4~GHz image, in Galactic
coordinates, of the candidate SNR \SNR(38.83)-(0.01) reported in
\citet{2017A&A...605A..58A}. Contour levels are $\pm 0.8, 1.6, 2.4, 3.2,
4.0$ mJy beam$^{-1}$ (with the negative contours dashed). This image is
observations made with the Very Large Array (VLA), in B, C and D
configurations, with a resolution of approximately
6~arcsec.}\label{f:double}
\end{figure}

\subsection{Recently proposed SNRs}\label{s:pp}

As noted above, the detailed version of the catalogue includes notes on
many objects that have been reported in the literature as possible or
probable SNRs. Here I discuss some of the recently proposed SNR
candidates in more detail.
\begin{itemize}
\item \citet{2015MNRAS.453.2082D} suggest that a region of radio
emission, which they label NGC 6334D, might be a SNR. This region, near
$l=351\fdeg6$, $b=0\fdeg2$, was identified from their 31-GHz observations
(with a resolution of $\sim 4.5$~arcmin), apparently with a non-thermal
radio spectrum. However, other available observations of this region do
not support a SNR identification for NGC 6334D.
\citeauthor{2015MNRAS.453.2082D} noted there are two sources in the
NRAO VLA Sky Survey (NVSS, \citealt{1998AJ....115.1693C}, at 1.4~GHz
with a resolution of 45 arcsec) in the region of NGC 6334D, with peaks
at 2.1 and 2.0 Jy beam$^{-1}$. Each of these sources have integrated
flux densities of about 3.8 Jy in the NVSS catalogue, and other
observations (e.g.\ \citealt{2007MNRAS.382..382M}) show they have
relatively flat radio spectra. They are each associated with one or more
compact {\sc H\,ii} regions identified by \citet{2005AJ....129..348G},
from higher resolution 5-GHz and IR observations. The NVSS sources are
separated by about 4 arcmin, and -- with flat radio spectra -- explain
the emission of NGC 6334D seen in
\citeauthor{2015MNRAS.453.2082D}'s lower resolution 31-GHz image. Higher
quality 1.4-GHz observations from the SGPS \citep{2006ApJS..167..230H}
do not show any obvious emission in this region -- apart from that from
the NVSS sources -- that might indicate a SNR.
\item A sample of `giant radio sources' identified in the NVSS from
pattern recognition techniques is presented by
\citet{2016ApJS..224...18P}. One of these sources, NVGRC
J205051.1$+$312728 -- which is annotated with `SNR?' as one of several
possibilities -- is actually part of the Cygnus Loop
($=$\SNR(74.0)-(8.5), e.g.\ \citealt{1990AJ....100.1927G}). Several
other of these sources also correspond to known SNRs, including other
parts of the Cygnus Loop.
\item As noted above, \citet{2017A&A...605A..58A} identify many SNR
candidates, which include several very small objects, 6 having a radius
of $\le 1$~arcmin. If these were SNRs, then they would have to be
physically small, even if in the distant Galaxy, and so would be
scientifically interesting. Higher resolution radio images are available
from the Multi-Array Galactic Plane Imaging Survey (MAGPIS,
\citealt{2006AJ....131.2525H})\footnote{see also: {\tt
https://third.ucllnl.org/gps/}} for several of these, but none of these
look like young SNRs. In particular \SNR(38.83)-(0.01) -- which was
reported as having a radius of 0.6~arcmin and an flux density at 1.4~GHz
of 10~mJy -- is clearly resolved into two compact sources, see
Fig.~\ref{f:double}, so is not a young SNR.
\item \citet{2018ApJ...866..100D} present observations of a small (only
about 15 arcsec in extent) radio shell, which they suggest may be a SNR.
If this was a SNR, it would have to be very young, given its small
angular size, even younger than the youngest known Galactic SNR
\SNR(1.9)+(0.3) (e.g.\ \citealt{ 2008MNRAS.387L..54G,
2008ApJ...680L..41R}). However, this source has already been identified
as a candidate PN by \citet{2015MNRAS.454.2586F}.
%
%
\end{itemize}

\section*{Acknowledgements}

I am grateful to colleagues for numerous comments on, and corrections
to the various versions of the Galactic SNR catalogue. This research
has made use of NASA's Astrophysics Data System Bibliographic Services,
and the SIMBAD database, operated at CDS, Strasbourg, France.

%
%

\onecolumn\relax
%
%
\input snrcat-table.tex

\input snrcat-table.txt

\endsnrcat

\end{document}

%% file: snrcat-table.tex
\newcount\linesdone
\global\linesdone=0
\newcount\processed
\global\processed=0
%
%
%
\makeatletter
\def\tablefont{\@setfontsize\tablefontsize{10.0}{11.0}}
\makeatother
%
%
\def\captiontext{294 Galactic supernova remnants: summary
  data.\labeltable{t:snrcat}}
\def\tops{%
  \setbox0=\vbox\bgroup\tablefont%
  \centerline{\small{\bf Table~1}. \captiontext}
  \centerline{\hrulefill}
  \vskip-2pt
  \centerline{%
    \hbox to 0.06\hsize{\hfil$l$\enskip}%
    \hbox to 0.065\hsize{\hfil$b$\enskip}%
    \hbox to 0.195\hsize{\hss\quad RA (J2000) Dec\hss}%
    \hbox to 0.12\hsize{\hfil size\hfil}%
    \hbox to 0.05\hsize{type\hfil}%
    \hbox to 0.085\hsize{\quad Flux at\hfil}%
    \hbox to 0.10\hsize{\hfil spectral\hfil}%
    \hbox to 0.329\hsize{\enspace other \hfil}%
    \hfill
  }
  \centerline{%
    \hbox to 0.06\hsize{\hfil/${}^\circ$\kern3pt}%
    \hbox to 0.065\hsize{\hfil/${}^\circ$\kern3pt}%
    \hbox to 0.115\hsize{\hfil/$({\rm h}$\enskip${\rm m}$\enskip${\rm s})$}%
    \hbox to 0.080\hsize{\hfil/$({}^\circ$\kern6pt${}'$)}%
    \hbox to 0.12\hsize{\hfil/arcmin\hfil}%
    \hbox to 0.05\hsize{}%
    \hbox to 0.085\hsize{\ 1~GHz/Jy\hss}%
    \hbox to 0.10\hsize{\hfil index\hfil}%
    \hbox to 0.329\hsize{\enspace name(s)\hfil}%
    \hfill
  }
  \vskip-6pt
  \centerline{\hrulefill} %
  \egroup\box0
  \bgroup
  \tablefont
}
%
%
%
%
%
%
%
%
%
%
%
%
\def\LONGITUDE #1 {\def\ldegrees{#1}}
\def\LATITUDE #1 {\def\bdegrees{#1}}
\def\RAHMS #1 #2 #3 {\def\rahms{#1~#2~#3}}
\def\DECDM #1 #2 {\def\decdm{#1~#2}}
\def\SIZE #1 {\edef\size{#1}}            
\def\ALPHA #1 {\def\spectralindex{#1}}
\def\FLUX1GHZ #1 {\def\fluxGHz{#1}}
\def\TYPE #1 {\def\type{#1}}
\def\NAMES #1\par{\def\names{#1}\ifnum\linesdone=0\tops\fi%
  \global\advance\linesdone by 1
  \global\advance\processed by 1
  \centerline{%
    \hbox to 0.06\hsize{\hfil$\ldegrees$}%
    \hbox to 0.065\hsize{\hfil$\bdegrees$}%
    \hbox to 0.11\hsize{\hfil$\rahms$}%
    \hbox to 0.085\hsize{\hfil$\decdm$}%
    \hbox to 0.12\hsize{\FormatSize{\size}\hss}%
    \hbox to 0.05\hsize{\enspace\type\hfil}%
    \hbox to 0.085\hsize{\FormatFlux{\fluxGHz}\hss}%
    \hbox to 0.10\hsize{\quad\spectralindex\hfil}%
    \hbox to 0.325\hsize{\enspace\names\hfil}%
    \hfill
  }
  \vskip -1pt
  \setbox0=\vbox\bgroup\hfuzz=20pt
}
%
%
\def\NOTES{\par}
\def\RADIO{\par}
\def\XRAY{\par}
\def\OPTICAL{\par}
\def\DISTANCE{\par}
\def\POINT{\par}
\def\REFERENCE{\par}
%
%
\def\PM/{\pm}
\def\X/{*} 
\def\I/{{\small I}}
\def\II/{{\small II}}
\def\III/{{\small III}}
\def\V/{{\small V}}
\def\VV/{{\small X}}       
\def\XIV/{{\small XIV}}
\def\AD/{{\small AD}}
\def\Halpha/{{H$\alpha$}}
\def\HCOplus/{{HCO$^+$}}
\def\etal/{{et al.}}
\def\fdeg{^\circ\mkern-7mu.\mkern1mu}
\def\fmin{'\mkern-5mu.}
\def\fsec{''\mkern-7mu.\mkern1mu}
\let\eightpoint=\footnotesize
%
%
\def\DATE #1\par{
  \egroup
  \ifnum\linesdone=55
    \egroup
    \global\linesdone=0
    \global\processed=0
    \vskip-6pt
    \centerline{\hrulefill}
    \vfill\eject
    \def\captiontext{(continued).}
  \fi
  \ifnum\processed=5
    \vskip 5pt plus 1pt minus 1pt
    \global\processed=0
  \fi
}
\newif\ifQuestion
\newif\ifDecimal
\def\SplitAtDecimal#1.#2\relax{\gdef\BeforeDecimal{#1}\gdef\AfterDecimal{#2}}
\def\EndDecimalTest{}
\def\SplitDecimal#1{\let\next\SplitAtDecimal%
  \expandafter\next#1.\relax%
  \ifx\AfterDecimal\EndDecimalTest%
    \Decimalfalse%
  \else%
    \Decimaltrue%
    \expandafter\next#1\relax%
  \fi}
\def\SplitAtQuestion#1?#2\relax{\gdef\BeforeQuestion{#1}\gdef\AfterQuestion{#2}}
\def\EndQuestionTest{}
\def\SplitQuestion#1{\let\next\SplitAtQuestion%
  \expandafter\next#1?\relax%
  \ifx\AfterQuestion\EndQuestionTest%
    \Questionfalse%
  \else%
    \Questiontrue%
    \expandafter\next#1\relax%
  \fi}
\def\FormatFlux#1{\Decimalfalse\Questionfalse\SplitDecimal{#1}%
   \ifDecimal%
     \setbox0=\hbox{\hbox to 0.045\hsize{\hfill\BeforeDecimal}\hbox
       to 0.040\hsize{.\AfterDecimal\hfill}}%
   \else%
     \SplitQuestion{#1}%
     \ifQuestion%
       \ifx\BeforeQuestion\EndQuestionTest%
         \setbox0=\hbox{\hbox to 0.045\hsize{\hfill?}\hbox
           to 0.040\hsize{\hfill}}%
       \else%
         \setbox0=\hbox{\hbox to 0.045\hsize{\hfill\BeforeQuestion}\hbox
           to 0.040\hsize{?\hfill}}%
        \fi%
     \else%
       \setbox0=\hbox{\hbox to 0.045\hsize{\hfill#1}\hbox
         to 0.040\hsize{\hfill}}%
     \fi%
   \fi\box0}
%
%
\newif\ifStar
\def\SplitAtStar#1*#2\relax{\gdef\BeforeStar{#1}\gdef\AfterStar{#2}}
\def\EndStarTest{}
\def\SplitStar#1{\let\next\SplitAtStar%
  \expandafter\next#1*\relax%
  \ifx\AfterStar\EndStarTest%
    \Starfalse%
  \else%
    \Startrue%
    \expandafter\next#1\relax%
  \fi}
\def\FormatSize#1{\Starfalse\Decimalfalse\Questionfalse\SplitStar{#1}%
   \ifStar%
     \setbox0=\hbox{\hbox
       to 0.065\hsize{\hfill\BeforeStar${\scriptstyle\times}$}\hbox
       to 0.055\hsize{\AfterStar\hfill}}%
   \else%
     \SplitDecimal{#1}%
     \ifDecimal%
       \setbox0=\hbox{\hbox to 0.07\hsize{\hfill\BeforeDecimal}\hbox
         to 0.05\hsize{.\AfterDecimal\hfill}}%
     \else%
       \SplitQuestion{#1}%
       \ifQuestion%
         \ifx\BeforeQuestion\EndQuestionTest%
           \setbox0=\hbox{\hbox to 0.07\hsize{\hfill?}\hbox
             to 0.05\hsize{\hfill}}%
         \else%
           \setbox0=\hbox{\hbox to 0.07\hsize{\hfill\BeforeQuestion}\hbox
             to 0.05\hsize{?\hfill}}%
          \fi%
       \else%
         \setbox0=\hbox{\hbox to 0.07\hsize{\hfill#1}\hbox
           to 0.05\hsize{\hfill}}%
       \fi%
     \fi%
   \fi\box0}
\def\endsnrcat{\egroup\vskip-6pt\centerline{\hrulefill}}

%% file: snrcat-table.txt
\LONGITUDE 0.0 \LATITUDE +0.0
\RAHMS 17 45 44 \DECDM -29 00
\SIZE 3.5\X/2.5 \TYPE S
\FLUX1GHZ 100? \ALPHA 0.8?
\NAMES Sgr A East

\DATE 9 Apr 2019

\LONGITUDE 0.3 \LATITUDE +0.0
\RAHMS 17 46 15 \DECDM -28 38
\SIZE 15\X/8 \TYPE S
\FLUX1GHZ 22 \ALPHA 0.6
\NAMES

\DATE 3 Apr 2019

\LONGITUDE 0.9 \LATITUDE +0.1
\RAHMS 17 47 21 \DECDM -28 09
\SIZE 8 \TYPE C
\FLUX1GHZ 18? \ALPHA varies
\NAMES

\DATE 15 Apr 2019

\LONGITUDE 1.0 \LATITUDE -0.1
\RAHMS 17 48 30 \DECDM -28 09
\SIZE 8 \TYPE S
\FLUX1GHZ 15 \ALPHA 0.6?
\NAMES

\DATE 4 Jun 2017

\LONGITUDE 1.4 \LATITUDE -0.1
\RAHMS 17 49 39 \DECDM -27 46
\SIZE 10 \TYPE S
\FLUX1GHZ 2? \ALPHA ?
\NAMES

\DATE 4 Jun 2017

\LONGITUDE 1.9 \LATITUDE +0.3
\RAHMS 17 48 45 \DECDM -27 10
\SIZE 1.5 \TYPE S
\FLUX1GHZ 0.6 \ALPHA 0.6
\NAMES

\DATE 3 Apr 2019

\LONGITUDE 3.7 \LATITUDE -0.2
\RAHMS 17 55 26 \DECDM -25 50
\SIZE 14\X/11 \TYPE S
\FLUX1GHZ 2.3 \ALPHA 0.65
\NAMES

\DATE 31 Mar 2006

\LONGITUDE 3.8 \LATITUDE +0.3
\RAHMS 17 52 55 \DECDM -25 28
\SIZE 18 \TYPE S?
\FLUX1GHZ 3? \ALPHA 0.6
\NAMES

\DATE 11 Jan 2004

\LONGITUDE 4.2 \LATITUDE -3.5
\RAHMS 18 08 55 \DECDM -27 03
\SIZE 28 \TYPE S
\FLUX1GHZ 3.2? \ALPHA 0.6?
\NAMES

\DATE 25 Apr 2014

\LONGITUDE 4.5 \LATITUDE +6.8
\RAHMS 17 30 42 \DECDM -21 29
\SIZE 3 \TYPE S
\FLUX1GHZ 19 \ALPHA 0.64
\NAMES Kepler, SN1604, 3C358

\DATE 9 Apr 2019

\LONGITUDE 4.8 \LATITUDE +6.2
\RAHMS 17 33 25 \DECDM -21 34
\SIZE 18 \TYPE S
\FLUX1GHZ 3 \ALPHA 0.6
\NAMES

\DATE 25 Apr 2014

\LONGITUDE 5.2 \LATITUDE -2.6
\RAHMS 18 07 30 \DECDM -25 45
\SIZE 18 \TYPE S
\FLUX1GHZ 2.6? \ALPHA 0.6?
\NAMES

\DATE 25 Apr 2014

\LONGITUDE 5.4 \LATITUDE -1.2
\RAHMS 18 02 10 \DECDM -24 54
\SIZE 35 \TYPE C?
\FLUX1GHZ 35? \ALPHA 0.2?
\NAMES Milne 56

\DATE 4 May 2017

\LONGITUDE 5.5 \LATITUDE +0.3
\RAHMS 17 57 04 \DECDM -24 00
\SIZE 15\X/12 \TYPE S
\FLUX1GHZ 5.5 \ALPHA 0.7
\NAMES

\DATE 11 May 2017

\LONGITUDE 5.9 \LATITUDE +3.1
\RAHMS 17 47 20 \DECDM -22 16
\SIZE 20 \TYPE S
\FLUX1GHZ 3.3? \ALPHA 0.4?
\NAMES

\DATE 25 Apr 2014

\LONGITUDE 6.1 \LATITUDE +0.5
\RAHMS 17 57 29 \DECDM -23 25
\SIZE 18\X/12 \TYPE S
\FLUX1GHZ 4.5 \ALPHA 0.9
\NAMES

\DATE 11 May 2017

\LONGITUDE 6.1 \LATITUDE +1.2
\RAHMS 17 54 55 \DECDM -23 05
\SIZE 30\X/26 \TYPE F
\FLUX1GHZ 4.0? \ALPHA 0.3?
\NAMES

\DATE 25 Apr 2014

\LONGITUDE 6.4 \LATITUDE -0.1
\RAHMS 18 00 30 \DECDM -23 26
\SIZE 48 \TYPE C
\FLUX1GHZ 310 \ALPHA varies
\NAMES W28

\DATE 15 Apr 2019

\LONGITUDE 6.4 \LATITUDE +4.0
\RAHMS 17 45 10 \DECDM -21 22
\SIZE 31 \TYPE S
\FLUX1GHZ 1.3? \ALPHA 0.4?
\NAMES

\DATE 25 Apr 2014

\LONGITUDE 6.5 \LATITUDE -0.4
\RAHMS 18 02 11 \DECDM -23 34
\SIZE 18 \TYPE S
\FLUX1GHZ 27 \ALPHA 0.6
\NAMES

\DATE 4 Jun 2017

\LONGITUDE 7.0 \LATITUDE -0.1
\RAHMS 18 01 50 \DECDM -22 54
\SIZE 15 \TYPE S
\FLUX1GHZ 2.5? \ALPHA 0.5?
\NAMES

\DATE 10 Oct 2001

\LONGITUDE 7.2 \LATITUDE +0.2
\RAHMS 18 01 07 \DECDM -22 38
\SIZE 12 \TYPE S
\FLUX1GHZ 2.8 \ALPHA 0.6
\NAMES

\DATE 25 Apr 2014

\LONGITUDE 7.7 \LATITUDE -3.7
\RAHMS 18 17 25 \DECDM -24 04
\SIZE 22 \TYPE S
\FLUX1GHZ 11 \ALPHA 0.32
\NAMES 1814$-$24

\DATE 9 Apr 2019

\LONGITUDE 8.3 \LATITUDE -0.0
\RAHMS 18 04 34 \DECDM -21 49
\SIZE 5\X/4 \TYPE S
\FLUX1GHZ 1.2 \ALPHA 0.6
\NAMES

\DATE 13 May 2017

\LONGITUDE 8.7 \LATITUDE -5.0
\RAHMS 18 24 10 \DECDM -23 48
\SIZE 26 \TYPE S
\FLUX1GHZ 4.4 \ALPHA 0.3
\NAMES

\DATE 3 Apr 2019

\LONGITUDE 8.7 \LATITUDE -0.1
\RAHMS 18 05 30 \DECDM -21 26
\SIZE 45 \TYPE S?
\FLUX1GHZ 80 \ALPHA 0.5
\NAMES (W30)

\DATE 13 May 2017

\LONGITUDE 8.9 \LATITUDE +0.4
\RAHMS 18 03 58 \DECDM -21 03
\SIZE 24 \TYPE S
\FLUX1GHZ 9 \ALPHA 0.6
\NAMES

\DATE 25 Apr 2014

\LONGITUDE 9.7 \LATITUDE -0.0
\RAHMS 18 07 22 \DECDM -20 35
\SIZE 15\X/11 \TYPE S
\FLUX1GHZ 3.7 \ALPHA 0.6
\NAMES

\DATE 13 May 2017

\LONGITUDE 9.8 \LATITUDE +0.6
\RAHMS 18 05 08 \DECDM -20 14
\SIZE 12 \TYPE S
\FLUX1GHZ 3.9 \ALPHA 0.5
\NAMES

\DATE 13 Aug 1998

\LONGITUDE 9.9 \LATITUDE -0.8
\RAHMS 18 10 41 \DECDM -20 43
\SIZE 12 \TYPE S
\FLUX1GHZ 6.7 \ALPHA 0.4
\NAMES

\DATE 13 May 2017

\LONGITUDE 10.5 \LATITUDE -0.0
\RAHMS 18 09 08 \DECDM -19 47
\SIZE 6 \TYPE S
\FLUX1GHZ 0.9 \ALPHA 0.6
\NAMES

\DATE 28 Apr 2017

\LONGITUDE 11.0 \LATITUDE -0.0
\RAHMS 18 10 04 \DECDM -19 25
\SIZE 11\X/9 \TYPE S
\FLUX1GHZ 1.3 \ALPHA 0.6
\NAMES

\DATE 9 Apr 2019

\LONGITUDE 11.1 \LATITUDE -1.0
\RAHMS 18 14 03 \DECDM -19 46
\SIZE 18\X/12 \TYPE S
\FLUX1GHZ 5.8 \ALPHA 0.5
\NAMES

\DATE 28 Apr 2014

\LONGITUDE 11.1 \LATITUDE -0.7
\RAHMS 18 12 46 \DECDM -19 38
\SIZE 11\X/7 \TYPE S
\FLUX1GHZ 1.0 \ALPHA 0.7
\NAMES

\DATE 29 Mar 2006

\LONGITUDE 11.1 \LATITUDE +0.1
\RAHMS 18 09 47 \DECDM -19 12
\SIZE 12\X/10 \TYPE S
\FLUX1GHZ 2.3 \ALPHA 0.4
\NAMES

\DATE 13 May 2017

\LONGITUDE 11.2 \LATITUDE -0.3
\RAHMS 18 11 27 \DECDM -19 25
\SIZE 4 \TYPE C
\FLUX1GHZ 22 \ALPHA 0.5
\NAMES

\DATE 16 Jun 2017

\LONGITUDE 11.4 \LATITUDE -0.1
\RAHMS 18 10 47 \DECDM -19 05
\SIZE 8 \TYPE S?
\FLUX1GHZ 6 \ALPHA 0.5
\NAMES

\DATE 13 May 2017

\LONGITUDE 11.8 \LATITUDE -0.2
\RAHMS 18 12 25 \DECDM -18 44
\SIZE 4 \TYPE S
\FLUX1GHZ 0.7 \ALPHA 0.3
\NAMES

\DATE 28 Apr 2017

\LONGITUDE 12.0 \LATITUDE -0.1
\RAHMS 18 12 11 \DECDM -18 37
\SIZE 7? \TYPE ?
\FLUX1GHZ 3.5 \ALPHA 0.7
\NAMES

\DATE 4 May 2017

\LONGITUDE 12.2 \LATITUDE +0.3
\RAHMS 18 11 17 \DECDM -18 10
\SIZE 6\X/5 \TYPE S
\FLUX1GHZ 0.8 \ALPHA 0.7
\NAMES

\DATE 13 May 2017

\LONGITUDE 12.5 \LATITUDE +0.2
\RAHMS 18 12 14 \DECDM -17 55
\SIZE 6\X/5 \TYPE C?
\FLUX1GHZ 0.6 \ALPHA 0.4
\NAMES

\DATE 25 Apr 2014

\LONGITUDE 12.7 \LATITUDE -0.0
\RAHMS 18 13 19 \DECDM -17 54
\SIZE 6 \TYPE S
\FLUX1GHZ 0.8 \ALPHA 0.8
\NAMES

\DATE 25 Apr 2014

\LONGITUDE 12.8 \LATITUDE -0.0
\RAHMS 18 13 37 \DECDM -17 49
\SIZE 3 \TYPE C?
\FLUX1GHZ 0.8 \ALPHA 0.5
\NAMES

\DATE 9 Apr 2019

\LONGITUDE 13.3 \LATITUDE -1.3
\RAHMS 18 19 20 \DECDM -18 00
\SIZE 70\X/40 \TYPE S?
\FLUX1GHZ ? \ALPHA ?
\NAMES

\DATE 1 Aug 2000

\LONGITUDE 13.5 \LATITUDE +0.2
\RAHMS 18 14 14 \DECDM -17 12
\SIZE 5\X/4 \TYPE S
\FLUX1GHZ 3.5? \ALPHA 1.0?
\NAMES

\DATE 11 May 2017

\LONGITUDE 14.1 \LATITUDE -0.1
\RAHMS 18 16 40 \DECDM -16 41
\SIZE 6\X/5 \TYPE S
\FLUX1GHZ 0.5 \ALPHA 0.6
\NAMES

\DATE 25 Apr 2014

\LONGITUDE 14.3 \LATITUDE +0.1
\RAHMS 18 15 58 \DECDM -16 27
\SIZE 5\X/4 \TYPE S
\FLUX1GHZ 0.6 \ALPHA 0.4
\NAMES

\DATE 20 May 2014

\LONGITUDE 15.1 \LATITUDE -1.6
\RAHMS 18 24 00 \DECDM -16 34
\SIZE 30\X/24 \TYPE S?
\FLUX1GHZ 5.5? \ALPHA 0.0?
\NAMES

\DATE 24 May 2014

\LONGITUDE 15.4 \LATITUDE +0.1
\RAHMS 18 18 02 \DECDM -15 27
\SIZE 15\X/14 \TYPE C?
\FLUX1GHZ 5.6 \ALPHA 0.62
\NAMES

\DATE 15 Apr 2019

\LONGITUDE 15.9 \LATITUDE +0.2
\RAHMS 18 18 52 \DECDM -15 02
\SIZE 7\X/5 \TYPE S?
\FLUX1GHZ 5.0 \ALPHA 0.63
\NAMES

\DATE 7 Jun 2019

\LONGITUDE 16.0 \LATITUDE -0.5
\RAHMS 18 21 56 \DECDM -15 14
\SIZE 15\X/10 \TYPE S
\FLUX1GHZ 2.7 \ALPHA 0.6
\NAMES

\DATE 11 May 2017

\LONGITUDE 16.2 \LATITUDE -2.7
\RAHMS 18 29 40 \DECDM -16 08
\SIZE 17 \TYPE S
\FLUX1GHZ 2.5 \ALPHA 0.4
\NAMES

\DATE 24 May 2014

\LONGITUDE 16.4 \LATITUDE -0.5
\RAHMS 18 22 38 \DECDM -14 55
\SIZE 13 \TYPE S
\FLUX1GHZ 4.6 \ALPHA 0.3?
\NAMES

\DATE 25 Apr 2014

\LONGITUDE 16.7 \LATITUDE +0.1
\RAHMS 18 20 56 \DECDM -14 20
\SIZE 4 \TYPE C
\FLUX1GHZ 3.0 \ALPHA 0.6
\NAMES

\DATE 9 Apr 2019

\LONGITUDE 17.0 \LATITUDE -0.0
\RAHMS 18 21 57 \DECDM -14 08
\SIZE 5 \TYPE S
\FLUX1GHZ 0.5 \ALPHA 0.5
\NAMES

\DATE 25 Apr 2014

\LONGITUDE 17.4 \LATITUDE -2.3
\RAHMS 18 30 55 \DECDM -14 52
\SIZE 24? \TYPE S
\FLUX1GHZ 5 \ALPHA 0.5?
\NAMES

\DATE 24 May 2014

\LONGITUDE 17.4 \LATITUDE -0.1
\RAHMS 18 23 08 \DECDM -13 46
\SIZE 6 \TYPE S
\FLUX1GHZ 0.4 \ALPHA 0.7
\NAMES

\DATE 13 May 2017

\LONGITUDE 17.8 \LATITUDE -2.6
\RAHMS 18 32 50 \DECDM -14 39
\SIZE 24 \TYPE S
\FLUX1GHZ 5 \ALPHA 0.5
\NAMES

\DATE 24 May 2014

\LONGITUDE 18.1 \LATITUDE -0.1
\RAHMS 18 24 34 \DECDM -13 11
\SIZE 8 \TYPE S
\FLUX1GHZ 4.6 \ALPHA 0.5
\NAMES

\DATE 9 Apr 2019

\LONGITUDE 18.6 \LATITUDE -0.2
\RAHMS 18 25 55 \DECDM -12 50
\SIZE 6 \TYPE S
\FLUX1GHZ 1.4 \ALPHA 0.4
\NAMES

\DATE 9 Apr 2019

\LONGITUDE 18.8 \LATITUDE +0.3
\RAHMS 18 23 58 \DECDM -12 23
\SIZE 17\X/11 \TYPE S
\FLUX1GHZ 33 \ALPHA 0.46
\NAMES Kes 67

\DATE 9 Apr 2019

\LONGITUDE 18.9 \LATITUDE -1.1
\RAHMS 18 29 50 \DECDM -12 58
\SIZE 33 \TYPE C?
\FLUX1GHZ 37 \ALPHA 0.39
\NAMES

\DATE 9 Apr 2019

\LONGITUDE 19.1 \LATITUDE +0.2
\RAHMS 18 24 56 \DECDM -12 07
\SIZE 27 \TYPE S
\FLUX1GHZ 10 \ALPHA 0.5
\NAMES

\DATE 29 Mar 2006

\LONGITUDE 20.0 \LATITUDE -0.2
\RAHMS 18 28 07 \DECDM -11 35
\SIZE 10 \TYPE F
\FLUX1GHZ 10 \ALPHA 0.1
\NAMES

\DATE 9 Apr 2019

\LONGITUDE 21.0 \LATITUDE -0.4
\RAHMS 18 31 12 \DECDM -10 47
\SIZE 9\X/7 \TYPE S
\FLUX1GHZ 1.1 \ALPHA 0.6
\NAMES

\DATE 29 Mar 2006

\LONGITUDE 21.5 \LATITUDE -0.9
\RAHMS 18 33 33 \DECDM -10 35
\SIZE 5 \TYPE C
\FLUX1GHZ 7 \ALPHA varies
\NAMES

\DATE 9 Apr 2019

\LONGITUDE 21.6 \LATITUDE -0.8
\RAHMS 18 33 40 \DECDM -10 25
\SIZE 13 \TYPE S
\FLUX1GHZ 1.4 \ALPHA 0.5?
\NAMES

\DATE 11 May 2017

\LONGITUDE 21.8 \LATITUDE -0.6
\RAHMS 18 32 45 \DECDM -10 08
\SIZE 20 \TYPE S
\FLUX1GHZ 65 \ALPHA 0.56
\NAMES Kes 69

\DATE 9 Apr 2019

\LONGITUDE 22.7 \LATITUDE -0.2
\RAHMS 18 33 15 \DECDM -09 13
\SIZE 26 \TYPE S?
\FLUX1GHZ 33 \ALPHA 0.6
\NAMES

\DATE 9 Apr 2019

\LONGITUDE 23.3 \LATITUDE -0.3
\RAHMS 18 34 45 \DECDM -08 48
\SIZE 27 \TYPE S
\FLUX1GHZ 70 \ALPHA 0.5
\NAMES W41

\DATE 9 Apr 2019

\LONGITUDE 24.7 \LATITUDE -0.6
\RAHMS 18 38 43 \DECDM -07 32
\SIZE 15? \TYPE S?
\FLUX1GHZ 8 \ALPHA 0.5
\NAMES

\DATE 9 Apr 2019

\LONGITUDE 24.7 \LATITUDE +0.6
\RAHMS 18 34 10 \DECDM -07 05
\SIZE 30\X/15 \TYPE C?
\FLUX1GHZ 20? \ALPHA 0.2?
\NAMES

\DATE 13 May 2017

\LONGITUDE 25.1 \LATITUDE -2.3
\RAHMS 18 45 10 \DECDM -08 00
\SIZE 80\X/30? \TYPE S
\FLUX1GHZ 8 \ALPHA 0.5?
\NAMES

\DATE 30 Apr 2014

\LONGITUDE 27.4 \LATITUDE +0.0
\RAHMS 18 41 19 \DECDM -04 56
\SIZE 4 \TYPE S
\FLUX1GHZ 6 \ALPHA 0.68
\NAMES 4C$-$04.71

\DATE 9 Apr 2019

\LONGITUDE 27.8 \LATITUDE +0.6
\RAHMS 18 39 50 \DECDM -04 24
\SIZE 50\X/30 \TYPE F
\FLUX1GHZ 30 \ALPHA varies
\NAMES

\DATE 11 May 2017

\LONGITUDE 28.6 \LATITUDE -0.1
\RAHMS 18 43 55 \DECDM -03 53
\SIZE 13\X/9 \TYPE S
\FLUX1GHZ 3? \ALPHA ?
\NAMES

\DATE 9 Apr 2019

\LONGITUDE 28.8 \LATITUDE +1.5
\RAHMS 18 39 00 \DECDM -02 55
\SIZE 100? \TYPE S?
\FLUX1GHZ ? \ALPHA 0.4?
\NAMES

\DATE 11 May 2017

\LONGITUDE 29.6 \LATITUDE +0.1
\RAHMS 18 44 52 \DECDM -02 57
\SIZE 5 \TYPE S
\FLUX1GHZ 1.5? \ALPHA 0.5?
\NAMES

\DATE 13 May 2017

\LONGITUDE 29.7 \LATITUDE -0.3
\RAHMS 18 46 25 \DECDM -02 59
\SIZE 3 \TYPE C
\FLUX1GHZ 10 \ALPHA 0.63
\NAMES Kes 75

\DATE 9 Apr 2019

\LONGITUDE 30.7 \LATITUDE -2.0
\RAHMS 18 54 25 \DECDM -02 54
\SIZE 16 \TYPE ?
\FLUX1GHZ 0.5? \ALPHA 0.7?
\NAMES

\DATE 13 Aug 1998

\LONGITUDE 30.7 \LATITUDE +1.0
\RAHMS 18 44 00 \DECDM -01 32
\SIZE 24\X/18 \TYPE S?
\FLUX1GHZ 6 \ALPHA 0.4
\NAMES

\DATE 28 Apr 2014

\LONGITUDE 31.5 \LATITUDE -0.6
\RAHMS 18 51 10 \DECDM -01 31
\SIZE 18? \TYPE S?
\FLUX1GHZ 2? \ALPHA ?
\NAMES

\DATE 17 Oct 2001

\LONGITUDE 31.9 \LATITUDE +0.0
\RAHMS 18 49 25 \DECDM -00 55
\SIZE 7\X/5 \TYPE S
\FLUX1GHZ 25 \ALPHA varies
\NAMES 3C391

\DATE 4 Apr 2019

\LONGITUDE 32.0 \LATITUDE -4.9
\RAHMS 19 06 00 \DECDM -03 00
\SIZE 60? \TYPE S?
\FLUX1GHZ 22? \ALPHA 0.5?
\NAMES 3C396.1

\DATE 25 Feb 1988

\LONGITUDE 32.1 \LATITUDE -0.9
\RAHMS 18 53 10 \DECDM -01 08
\SIZE 40? \TYPE C?
\FLUX1GHZ ? \ALPHA ?
\NAMES

\DATE 11 May 2017

\LONGITUDE 32.4 \LATITUDE +0.1
\RAHMS 18 50 05 \DECDM -00 25
\SIZE 6 \TYPE S
\FLUX1GHZ 0.25? \ALPHA ?
\NAMES

\DATE 13 May 2017

\LONGITUDE 32.8 \LATITUDE -0.1
\RAHMS 18 51 25 \DECDM -00 08
\SIZE 22\X/15 \TYPE S?
\FLUX1GHZ 11? \ALPHA 0.2?
\NAMES Kes 78

\DATE 9 Apr 2019

\LONGITUDE 33.2 \LATITUDE -0.6
\RAHMS 18 53 50 \DECDM -00 02
\SIZE 18 \TYPE S
\FLUX1GHZ 3.5 \ALPHA varies
\NAMES

\DATE 11 May 2017

\LONGITUDE 33.6 \LATITUDE +0.1
\RAHMS 18 52 48 \DECDM +00 41
\SIZE 10 \TYPE S
\FLUX1GHZ 20 \ALPHA 0.51
\NAMES Kes 79, 4C00.70, HC13

\DATE 9 Apr 2019

\LONGITUDE 34.7 \LATITUDE -0.4
\RAHMS 18 56 00 \DECDM +01 22
\SIZE 35\X/27 \TYPE C
\FLUX1GHZ 240 \ALPHA 0.37
\NAMES W44, 3C392

\DATE 9 Apr 2019

\LONGITUDE 35.6 \LATITUDE -0.4
\RAHMS 18 57 55 \DECDM +02 13
\SIZE 15\X/11 \TYPE S?
\FLUX1GHZ 9 \ALPHA 0.5
\NAMES

\DATE 9 Apr 2019

\LONGITUDE 36.6 \LATITUDE -0.7
\RAHMS 19 00 35 \DECDM +02 56
\SIZE 25? \TYPE S?
\FLUX1GHZ 1.0 \ALPHA 0.7?
\NAMES

\DATE 24 May 2014

\LONGITUDE 36.6 \LATITUDE +2.6
\RAHMS 18 48 49 \DECDM +04 26
\SIZE 17\X/13? \TYPE S
\FLUX1GHZ 0.7? \ALPHA 0.5?
\NAMES

\DATE 19 May 1992

\LONGITUDE 38.7 \LATITUDE -1.3
\RAHMS 19 06 40 \DECDM +04 28
\SIZE 32\X/19? \TYPE S
\FLUX1GHZ ? \ALPHA ?
\NAMES

\DATE 5 Jun 2019

\LONGITUDE 39.2 \LATITUDE -0.3
\RAHMS 19 04 08 \DECDM +05 28
\SIZE 8\X/6 \TYPE C
\FLUX1GHZ 18 \ALPHA 0.34
\NAMES 3C396, HC24, NRAO 593

\DATE 9 Apr 2019

\LONGITUDE 39.7 \LATITUDE -2.0
\RAHMS 19 12 20 \DECDM +04 55
\SIZE 120\X/60 \TYPE ?
\FLUX1GHZ 85? \ALPHA 0.7?
\NAMES W50, SS433

\DATE 9 Apr 2019

\LONGITUDE 40.5 \LATITUDE -0.5
\RAHMS 19 07 10 \DECDM +06 31
\SIZE 22 \TYPE S
\FLUX1GHZ 11 \ALPHA 0.4
\NAMES

\DATE 9 Apr 2019

\LONGITUDE 41.1 \LATITUDE -0.3
\RAHMS 19 07 34 \DECDM +07 08
\SIZE 4.5\X/2.5 \TYPE S
\FLUX1GHZ 25 \ALPHA 0.50
\NAMES 3C397

\DATE 9 Apr 2019

\LONGITUDE 41.5 \LATITUDE +0.4
\RAHMS 19 05 50 \DECDM +07 46
\SIZE 10 \TYPE S?
\FLUX1GHZ 1? \ALPHA ?
\NAMES

\DATE 15 May 2014

\LONGITUDE 42.0 \LATITUDE -0.1
\RAHMS 19 08 10 \DECDM +08 00
\SIZE 8 \TYPE S?
\FLUX1GHZ 0.5? \ALPHA ?
\NAMES

\DATE 16 May 2014

\LONGITUDE 42.8 \LATITUDE +0.6
\RAHMS 19 07 20 \DECDM +09 05
\SIZE 24 \TYPE S
\FLUX1GHZ 3? \ALPHA 0.5?
\NAMES

\DATE 11 Jan 2004

\LONGITUDE 43.3 \LATITUDE -0.2
\RAHMS 19 11 08 \DECDM +09 06
\SIZE 4\X/3 \TYPE S
\FLUX1GHZ 38 \ALPHA 0.46
\NAMES W49B

\DATE 15 Apr 2019

\LONGITUDE 43.9 \LATITUDE +1.6
\RAHMS 19 05 50 \DECDM +10 30
\SIZE 60? \TYPE S?
\FLUX1GHZ 9.0 \ALPHA 0.5
\NAMES

\DATE 24 May 2014

\LONGITUDE 45.7 \LATITUDE -0.4
\RAHMS 19 16 25 \DECDM +11 09
\SIZE 22 \TYPE S
\FLUX1GHZ 4.2? \ALPHA 0.4?
\NAMES

\DATE 13 May 2017

\LONGITUDE 46.8 \LATITUDE -0.3
\RAHMS 19 18 10 \DECDM +12 09
\SIZE 15 \TYPE S
\FLUX1GHZ 17 \ALPHA 0.54
\NAMES (HC30)

\DATE 12 Jun 2019

\LONGITUDE 49.2 \LATITUDE -0.7
\RAHMS 19 23 50 \DECDM +14 06
\SIZE 30 \TYPE S?
\FLUX1GHZ 160? \ALPHA 0.3?
\NAMES (W51)

\DATE 9 Apr 2019

\LONGITUDE 53.4 \LATITUDE +0.0
\RAHMS 19 29 57 \DECDM +18 10
\SIZE 10? \TYPE S
\FLUX1GHZ 1.5 \ALPHA 0.6?
\NAMES

\DATE 10 Apr 2019

\LONGITUDE 53.6 \LATITUDE -2.2
\RAHMS 19 38 50 \DECDM +17 14
\SIZE 33\X/28 \TYPE S
\FLUX1GHZ 8 \ALPHA 0.50
\NAMES 3C400.2, NRAO 611

\DATE 9 Apr 2019

\LONGITUDE 54.1 \LATITUDE +0.3
\RAHMS 19 30 31 \DECDM +18 52
\SIZE 12? \TYPE C?
\FLUX1GHZ 0.5 \ALPHA 0.1
\NAMES

\DATE 11 Apr 2019

\LONGITUDE 54.4 \LATITUDE -0.3
\RAHMS 19 33 20 \DECDM +18 56
\SIZE 40 \TYPE S
\FLUX1GHZ 28 \ALPHA 0.5
\NAMES (HC40)

\DATE 9 Apr 2019

\LONGITUDE 55.0 \LATITUDE +0.3
\RAHMS 19 32 00 \DECDM +19 50
\SIZE 20\X/15? \TYPE S
\FLUX1GHZ 0.5? \ALPHA 0.5?
\NAMES

\DATE 9 Apr 2019

\LONGITUDE 55.7 \LATITUDE +3.4
\RAHMS 19 21 20 \DECDM +21 44
\SIZE 23 \TYPE S
\FLUX1GHZ 1? \ALPHA 0.3?
\NAMES

\DATE 24 May 2014

\LONGITUDE 57.2 \LATITUDE +0.8
\RAHMS 19 34 59 \DECDM +21 57
\SIZE 12? \TYPE S?
\FLUX1GHZ 1.8 \ALPHA 0.35
\NAMES (4C21.53)

\DATE 9 Apr 2019

\LONGITUDE 59.5 \LATITUDE +0.1
\RAHMS 19 42 33 \DECDM +23 35
\SIZE 15 \TYPE S
\FLUX1GHZ 3? \ALPHA ?
\NAMES

\DATE 25 Apr 2014

\LONGITUDE 63.7 \LATITUDE +1.1
\RAHMS 19 47 52 \DECDM +27 45
\SIZE 8 \TYPE F
\FLUX1GHZ 1.8 \ALPHA 0.24
\NAMES

\DATE 9 Apr 2019

\LONGITUDE 64.5 \LATITUDE +0.9
\RAHMS 19 50 25 \DECDM +28 16
\SIZE 8 \TYPE S?
\FLUX1GHZ 0.15? \ALPHA 0.5
\NAMES

\DATE 9 Apr 2019

\LONGITUDE 65.1 \LATITUDE +0.6
\RAHMS 19 54 40 \DECDM +28 35
\SIZE 90\X/50 \TYPE S
\FLUX1GHZ 5.5 \ALPHA 0.61
\NAMES

\DATE 11 May 2017

\LONGITUDE 65.3 \LATITUDE +5.7
\RAHMS 19 33 00 \DECDM +31 10
\SIZE 310\X/240 \TYPE S?
\FLUX1GHZ 42 \ALPHA 0.6
\NAMES

\DATE 28 Apr 2014

\LONGITUDE 65.7 \LATITUDE +1.2
\RAHMS 19 52 10 \DECDM +29 26
\SIZE 22 \TYPE F
\FLUX1GHZ 5.1 \ALPHA varies
\NAMES DA 495

\DATE 11 May 2017

\LONGITUDE 66.0 \LATITUDE -0.0
\RAHMS 19 57 50 \DECDM +29 03
\SIZE 31\X/25? \TYPE S
\FLUX1GHZ ? \ALPHA ?
\NAMES

\DATE 9 Apr 2019

\LONGITUDE 67.6 \LATITUDE +0.9
\RAHMS 19 57 45 \DECDM +30 53
\SIZE 50\X/45? \TYPE S
\FLUX1GHZ ? \ALPHA ?
\NAMES

\DATE 9 Apr 2019

\LONGITUDE 67.7 \LATITUDE +1.8
\RAHMS 19 54 32 \DECDM +31 29
\SIZE 15\X/12 \TYPE S
\FLUX1GHZ 1.0 \ALPHA 0.61
\NAMES

\DATE 9 Apr 2019

\LONGITUDE 67.8 \LATITUDE +0.5
\RAHMS 20 00 00 \DECDM +30 51
\SIZE 7\X/5 \TYPE ?
\FLUX1GHZ ? \ALPHA ?
\NAMES

\DATE 30 Apr 2014

\LONGITUDE 68.6 \LATITUDE -1.2
\RAHMS 20 08 40 \DECDM +30 37
\SIZE 23 \TYPE ?
\FLUX1GHZ 1.1 \ALPHA 0.2
\NAMES

\DATE 24 May 2014

\LONGITUDE 69.0 \LATITUDE +2.7
\RAHMS 19 53 20 \DECDM +32 55
\SIZE 80? \TYPE ?
\FLUX1GHZ 120? \ALPHA varies
\NAMES CTB 80

\DATE 9 Apr 2019

\LONGITUDE 69.7 \LATITUDE +1.0
\RAHMS 20 02 40 \DECDM +32 43
\SIZE 16\X/14 \TYPE S
\FLUX1GHZ 2.0 \ALPHA 0.7
\NAMES

\DATE 28 Apr 2014

\LONGITUDE 70.0 \LATITUDE -21.5
\RAHMS 21 24 00 \DECDM +19 23
\SIZE 330\X/240 \TYPE S
\FLUX1GHZ ? \ALPHA ?
\NAMES

\DATE 11 Jun 2017

\LONGITUDE 73.9 \LATITUDE +0.9
\RAHMS 20 14 15 \DECDM +36 12
\SIZE 27 \TYPE S?
\FLUX1GHZ 9 \ALPHA 0.23
\NAMES

\DATE 13 May 2017

\LONGITUDE 74.0 \LATITUDE -8.5
\RAHMS 20 51 00 \DECDM +30 40
\SIZE 230\X/160 \TYPE S
\FLUX1GHZ 210 \ALPHA varies
\NAMES Cygnus Loop

\DATE 9 Apr 2019

\LONGITUDE 74.9 \LATITUDE +1.2
\RAHMS 20 16 02 \DECDM +37 12
\SIZE 8\X/6 \TYPE F
\FLUX1GHZ 9 \ALPHA varies
\NAMES CTB 87

\DATE 9 Apr 2019

\LONGITUDE 76.9 \LATITUDE +1.0
\RAHMS 20 22 20 \DECDM +38 43
\SIZE 9 \TYPE C
\FLUX1GHZ 2? \ALPHA ?
\NAMES

\DATE 9 Apr 2019

\LONGITUDE 78.2 \LATITUDE +2.1
\RAHMS 20 20 50 \DECDM +40 26
\SIZE 60 \TYPE S
\FLUX1GHZ 320 \ALPHA 0.51
\NAMES DR4, $\gamma$ Cygni SNR

\DATE 9 Apr 2019

\LONGITUDE 82.2 \LATITUDE +5.3
\RAHMS 20 19 00 \DECDM +45 30
\SIZE 95\X/65 \TYPE S
\FLUX1GHZ 120? \ALPHA 0.5?
\NAMES W63

\DATE 9 Apr 2019

\LONGITUDE 83.0 \LATITUDE -0.3
\RAHMS 20 46 55 \DECDM +42 52
\SIZE 9\X/7 \TYPE S
\FLUX1GHZ 1 \ALPHA 0.4
\NAMES

\DATE 9 Apr 2019

\LONGITUDE 84.2 \LATITUDE -0.8
\RAHMS 20 53 20 \DECDM +43 27
\SIZE 20\X/16 \TYPE S
\FLUX1GHZ 11 \ALPHA 0.5
\NAMES

\DATE 28 Apr 2014

\LONGITUDE 85.4 \LATITUDE +0.7
\RAHMS 20 50 40 \DECDM +45 22
\SIZE 24? \TYPE S
\FLUX1GHZ ? \ALPHA 0.2
\NAMES

\DATE 9 Apr 2019

\LONGITUDE 85.9 \LATITUDE -0.6
\RAHMS 20 58 40 \DECDM +44 53
\SIZE 24 \TYPE S
\FLUX1GHZ ? \ALPHA 0.2
\NAMES

\DATE 28 Apr 2014

\LONGITUDE 89.0 \LATITUDE +4.7
\RAHMS 20 45 00 \DECDM +50 35
\SIZE 120\X/90 \TYPE S
\FLUX1GHZ 220 \ALPHA 0.38
\NAMES HB21

\DATE 9 Apr 2019

\LONGITUDE 93.3 \LATITUDE +6.9
\RAHMS 20 52 25 \DECDM +55 21
\SIZE 27\X/20 \TYPE C?
\FLUX1GHZ 9 \ALPHA 0.45
\NAMES DA 530, 4C(T)55.38.1

\DATE 28 Apr 2014

\LONGITUDE 93.7 \LATITUDE -0.2
\RAHMS 21 29 20 \DECDM +50 50
\SIZE 80 \TYPE S
\FLUX1GHZ 65 \ALPHA 0.65
\NAMES CTB 104A, DA 551

\DATE 28 Apr 2014

\LONGITUDE 94.0 \LATITUDE +1.0
\RAHMS 21 24 50 \DECDM +51 53
\SIZE 30\X/25 \TYPE S
\FLUX1GHZ 13 \ALPHA 0.45
\NAMES 3C434.1

\DATE 24 May 2014

\LONGITUDE 96.0 \LATITUDE +2.0
\RAHMS 21 30 30 \DECDM +53 59
\SIZE 26 \TYPE S
\FLUX1GHZ 0.35 \ALPHA 0.6
\NAMES

\DATE 24 May 2014

\LONGITUDE 106.3 \LATITUDE +2.7
\RAHMS 22 27 30 \DECDM +60 50
\SIZE 60\X/24 \TYPE C?
\FLUX1GHZ 6 \ALPHA 0.6
\NAMES

\DATE 28 Apr 2017

\LONGITUDE 108.2 \LATITUDE -0.6
\RAHMS 22 53 40 \DECDM +58 50
\SIZE 70\X/54 \TYPE S
\FLUX1GHZ 8 \ALPHA 0.5
\NAMES

\DATE 16 Mar 2009

\LONGITUDE 109.1 \LATITUDE -1.0
\RAHMS 23 01 35 \DECDM +58 53
\SIZE 28 \TYPE S
\FLUX1GHZ 20 \ALPHA 0.45
\NAMES CTB 109

\DATE 5 Jun 2019

\LONGITUDE 111.7 \LATITUDE -2.1
\RAHMS 23 23 26 \DECDM +58 48
\SIZE 5 \TYPE S
\FLUX1GHZ 2300 \ALPHA 0.77
\NAMES Cassiopeia A, 3C461

\DATE 9 Apr 2019

\LONGITUDE 113.0 \LATITUDE +0.2
\RAHMS 23 26 50 \DECDM +61 26
\SIZE 40\X/17? \TYPE ?
\FLUX1GHZ 4 \ALPHA 0.5?
\NAMES

\DATE 28 Apr 2017

\LONGITUDE 114.3 \LATITUDE +0.3
\RAHMS 23 37 00 \DECDM +61 55
\SIZE 90\X/55 \TYPE S
\FLUX1GHZ 5.5 \ALPHA 0.5
\NAMES

\DATE 28 Apr 2014

\LONGITUDE 116.5 \LATITUDE +1.1
\RAHMS 23 53 40 \DECDM +63 15
\SIZE 80\X/60 \TYPE S
\FLUX1GHZ 10 \ALPHA 0.5
\NAMES

\DATE 28 Apr 2014

\LONGITUDE 116.9 \LATITUDE +0.2
\RAHMS 23 59 10 \DECDM +62 26
\SIZE 34 \TYPE S
\FLUX1GHZ 8 \ALPHA 0.57
\NAMES CTB 1

\DATE 11 Jun 2019

\LONGITUDE 119.5 \LATITUDE +10.2
\RAHMS 00 06 40 \DECDM +72 45
\SIZE 90? \TYPE S
\FLUX1GHZ 36 \ALPHA 0.6
\NAMES CTA 1

\DATE 9 Apr 2019

\LONGITUDE 120.1 \LATITUDE +1.4
\RAHMS 00 25 18 \DECDM +64 09
\SIZE 8 \TYPE S
\FLUX1GHZ 50 \ALPHA 0.58
\NAMES Tycho, 3C10, SN1572

\DATE 9 Apr 2019

\LONGITUDE 126.2 \LATITUDE +1.6
\RAHMS 01 22 00 \DECDM +64 15
\SIZE 70 \TYPE S?
\FLUX1GHZ 6 \ALPHA 0.5
\NAMES

\DATE 20 Feb 2009

\LONGITUDE 127.1 \LATITUDE +0.5
\RAHMS 01 28 20 \DECDM +63 10
\SIZE 45 \TYPE S
\FLUX1GHZ 12 \ALPHA 0.45
\NAMES R5

\DATE 5 May 2017

\LONGITUDE 130.7 \LATITUDE +3.1
\RAHMS 02 05 41 \DECDM +64 49
\SIZE 9\X/5 \TYPE F
\FLUX1GHZ 33 \ALPHA 0.07
\NAMES 3C58, SN1181

\DATE 13 May 2017

\LONGITUDE 132.7 \LATITUDE +1.3
\RAHMS 02 17 40 \DECDM +62 45
\SIZE 80 \TYPE S
\FLUX1GHZ 45 \ALPHA 0.6
\NAMES HB3

\DATE 13 May 2017

\LONGITUDE 150.3 \LATITUDE +4.5
\RAHMS 04 27 00 \DECDM +55 28
\SIZE 180\X/150 \TYPE S
\FLUX1GHZ ? \ALPHA ?
\NAMES

\DATE 10 Apr 2019

\LONGITUDE 152.4 \LATITUDE -2.1
\RAHMS 04 07 50 \DECDM +49 11
\SIZE 100\X/95 \TYPE S
\FLUX1GHZ 3.5? \ALPHA 0.7?
\NAMES

\DATE 30 Apr 2014

\LONGITUDE 156.2 \LATITUDE +5.7
\RAHMS 04 58 40 \DECDM +51 50
\SIZE 110 \TYPE S
\FLUX1GHZ 5 \ALPHA 0.5
\NAMES

\DATE 13 May 2017

\LONGITUDE 159.6 \LATITUDE +7.3
\RAHMS 05 20 00 \DECDM +50 00
\SIZE 240\X/180? \TYPE S
\FLUX1GHZ ? \ALPHA ?
\NAMES

\DATE 8 Jun 2019

\LONGITUDE 160.9 \LATITUDE +2.6
\RAHMS 05 01 00 \DECDM +46 40
\SIZE 140\X/120 \TYPE S
\FLUX1GHZ 110 \ALPHA 0.64
\NAMES HB9

\DATE 8 Jun 2019

\LONGITUDE 166.0 \LATITUDE +4.3
\RAHMS 05 26 30 \DECDM +42 56
\SIZE 55\X/35 \TYPE S
\FLUX1GHZ 7 \ALPHA 0.37
\NAMES VRO 42.05.01

\DATE 9 Apr 2019

\LONGITUDE 178.2 \LATITUDE -4.2
\RAHMS 05 25 05 \DECDM +28 11
\SIZE 72\X/62 \TYPE S
\FLUX1GHZ 2 \ALPHA 0.5
\NAMES

\DATE 28 Apr 2017

\LONGITUDE 179.0 \LATITUDE +2.6
\RAHMS 05 53 40 \DECDM +31 05
\SIZE 70 \TYPE S?
\FLUX1GHZ 7 \ALPHA 0.4
\NAMES

\DATE 8 Jun 2019

\LONGITUDE 180.0 \LATITUDE -1.7
\RAHMS 05 39 00 \DECDM +27 50
\SIZE 180 \TYPE S
\FLUX1GHZ 65 \ALPHA varies
\NAMES S147

\DATE 9 Apr 2019

\LONGITUDE 181.1 \LATITUDE +9.5
\RAHMS 06 26 40 \DECDM +32 30
\SIZE 74 \TYPE S
\FLUX1GHZ ? \ALPHA 0.45?
\NAMES

\DATE 3 Apr 2019

\LONGITUDE 182.4 \LATITUDE +4.3
\RAHMS 06 08 10 \DECDM +29 00
\SIZE 50 \TYPE S
\FLUX1GHZ 0.5 \ALPHA 0.4
\NAMES

\DATE 28 Apr 2014

\LONGITUDE 184.6 \LATITUDE -5.8
\RAHMS 05 34 31 \DECDM +22 01
\SIZE 7\X/5 \TYPE F
\FLUX1GHZ 900 \ALPHA 0.30
\NAMES Crab Nebula, 3C144, SN1054

\DATE 9 Apr 2019

\LONGITUDE 189.1 \LATITUDE +3.0
\RAHMS 06 17 00 \DECDM +22 34
\SIZE 45 \TYPE C
\FLUX1GHZ 165 \ALPHA 0.36
\NAMES IC443, 3C157

\DATE 9 Apr 2019

\LONGITUDE 190.9 \LATITUDE -2.2
\RAHMS 06 01 55 \DECDM +18 24
\SIZE 70\X/60 \TYPE S
\FLUX1GHZ 1.3? \ALPHA 0.7?
\NAMES

\DATE 20 May 2014

\LONGITUDE 205.5 \LATITUDE +0.5
\RAHMS 06 39 00 \DECDM +06 30
\SIZE 220 \TYPE S
\FLUX1GHZ 140 \ALPHA 0.4
\NAMES Monoceros Nebula

\DATE 9 Apr 2019

\LONGITUDE 206.9 \LATITUDE +2.3
\RAHMS 06 48 40 \DECDM +06 26
\SIZE 60\X/40 \TYPE S?
\FLUX1GHZ 6 \ALPHA 0.5
\NAMES PKS 0646$+$06

\DATE 9 Apr 2019

\LONGITUDE 213.0 \LATITUDE -0.6
\RAHMS 06 50 50 \DECDM -00 30
\SIZE 160\X/140? \TYPE S
\FLUX1GHZ 21 \ALPHA 0.4
\NAMES

\DATE 9 Apr 2019

\LONGITUDE 260.4 \LATITUDE -3.4
\RAHMS 08 22 10 \DECDM -43 00
\SIZE 60\X/50 \TYPE S
\FLUX1GHZ 130 \ALPHA 0.5
\NAMES Puppis A, MSH 08$-$4{\sl 4}

\DATE 9 Apr 2019

\LONGITUDE 261.9 \LATITUDE +5.5
\RAHMS 09 04 20 \DECDM -38 42
\SIZE 40\X/30 \TYPE S
\FLUX1GHZ 10? \ALPHA 0.4?
\NAMES

\DATE 13 Aug 1998

\LONGITUDE 263.9 \LATITUDE -3.3
\RAHMS 08 34 00 \DECDM -45 50
\SIZE 255 \TYPE C
\FLUX1GHZ 1750 \ALPHA varies
\NAMES Vela (XYZ)

\DATE 15 Apr 2019

\LONGITUDE 266.2 \LATITUDE -1.2
\RAHMS 08 52 00 \DECDM -46 20
\SIZE 120 \TYPE S
\FLUX1GHZ 50? \ALPHA 0.3?
\NAMES RX J0852.0$-$4622

\DATE 15 Apr 2019

\LONGITUDE 272.2 \LATITUDE -3.2
\RAHMS 09 06 50 \DECDM -52 07
\SIZE 15? \TYPE S?
\FLUX1GHZ 0.4 \ALPHA 0.6
\NAMES

\DATE 12 Jun 2017

\LONGITUDE 279.0 \LATITUDE +1.1
\RAHMS 09 57 40 \DECDM -53 15
\SIZE 95 \TYPE S
\FLUX1GHZ 30? \ALPHA 0.6?
\NAMES

\DATE 25 Apr 2014

\LONGITUDE 284.3 \LATITUDE -1.8
\RAHMS 10 18 15 \DECDM -59 00
\SIZE 24? \TYPE S
\FLUX1GHZ 11? \ALPHA 0.3?
\NAMES MSH 10$-$5{\sl 3}

\DATE 9 Apr 2019

\LONGITUDE 286.5 \LATITUDE -1.2
\RAHMS 10 35 40 \DECDM -59 42
\SIZE 26\X/6 \TYPE S?
\FLUX1GHZ 1.4? \ALPHA ?
\NAMES

\DATE 25 Apr 2014

\LONGITUDE 289.7 \LATITUDE -0.3
\RAHMS 11 01 15 \DECDM -60 18
\SIZE 18\X/14 \TYPE S
\FLUX1GHZ 6.2 \ALPHA 0.2?
\NAMES

\DATE 21 Aug 1996

\LONGITUDE 290.1 \LATITUDE -0.8
\RAHMS 11 03 05 \DECDM -60 56
\SIZE 19\X/14 \TYPE S
\FLUX1GHZ 42 \ALPHA 0.4
\NAMES MSH 11$-$6{\sl 1}A

\DATE 13 May 2017

\LONGITUDE 291.0 \LATITUDE -0.1
\RAHMS 11 11 54 \DECDM -60 38
\SIZE 15\X/13 \TYPE C
\FLUX1GHZ 16 \ALPHA 0.29
\NAMES (MSH 11$-$6{\sl 2})

\DATE 13 May 2017

\LONGITUDE 292.0 \LATITUDE +1.8
\RAHMS 11 24 36 \DECDM -59 16
\SIZE 12\X/8 \TYPE C
\FLUX1GHZ 15 \ALPHA 0.4
\NAMES MSH 11$-$5{\sl 4}

\DATE 16 Jun 2017

\LONGITUDE 292.2 \LATITUDE -0.5
\RAHMS 11 19 20 \DECDM -61 28
\SIZE 20\X/15 \TYPE S
\FLUX1GHZ 7 \ALPHA 0.5
\NAMES

\DATE 13 May 2017

\LONGITUDE 293.8 \LATITUDE +0.6
\RAHMS 11 35 00 \DECDM -60 54
\SIZE 20 \TYPE C
\FLUX1GHZ 5? \ALPHA 0.6?
\NAMES

\DATE 21 Aug 1996

\LONGITUDE 294.1 \LATITUDE -0.0
\RAHMS 11 36 10 \DECDM -61 38
\SIZE 40 \TYPE S
\FLUX1GHZ $>$2? \ALPHA ?
\NAMES

\DATE 21 Aug 1996

\LONGITUDE 296.1 \LATITUDE -0.5
\RAHMS 11 51 10 \DECDM -62 34
\SIZE 37\X/25 \TYPE S
\FLUX1GHZ 8? \ALPHA 0.6?
\NAMES

\DATE 24 Apr 2014

\LONGITUDE 296.5 \LATITUDE +10.0
\RAHMS 12 09 40 \DECDM -52 25
\SIZE 90\X/65 \TYPE S
\FLUX1GHZ 48 \ALPHA 0.5
\NAMES PKS 1209$-$51/52

\DATE 9 Apr 2019

\LONGITUDE 296.7 \LATITUDE -0.9
\RAHMS 11 55 30 \DECDM -63 08
\SIZE 15\X/8 \TYPE S
\FLUX1GHZ 3 \ALPHA 0.5
\NAMES

\DATE 12 Jun 2017

\LONGITUDE 296.8 \LATITUDE -0.3
\RAHMS 11 58 30 \DECDM -62 35
\SIZE 20\X/14 \TYPE S
\FLUX1GHZ 9 \ALPHA 0.6
\NAMES 1156$-$62

\DATE 24 Apr 2014

\LONGITUDE 298.5 \LATITUDE -0.3
\RAHMS 12 12 40 \DECDM -62 52
\SIZE 5? \TYPE ?
\FLUX1GHZ 5? \ALPHA 0.4?
\NAMES

\DATE 16 Mar 2009

\LONGITUDE 298.6 \LATITUDE -0.0
\RAHMS 12 13 41 \DECDM -62 37
\SIZE 12\X/9 \TYPE S
\FLUX1GHZ 5? \ALPHA 0.3
\NAMES

\DATE 13 May 2017

\LONGITUDE 299.2 \LATITUDE -2.9
\RAHMS 12 15 13 \DECDM -65 30
\SIZE 18\X/11 \TYPE S
\FLUX1GHZ 0.5? \ALPHA ?
\NAMES

\DATE 9 Apr 2019

\LONGITUDE 299.6 \LATITUDE -0.5
\RAHMS 12 21 45 \DECDM -63 09
\SIZE 13 \TYPE S
\FLUX1GHZ 1.0? \ALPHA ?
\NAMES

\DATE 21 Aug 1996

\LONGITUDE 301.4 \LATITUDE -1.0
\RAHMS 12 37 55 \DECDM -63 49
\SIZE 37\X/23 \TYPE S
\FLUX1GHZ 2.1? \ALPHA ?
\NAMES

\DATE 21 Aug 1996

\LONGITUDE 302.3 \LATITUDE +0.7
\RAHMS 12 45 55 \DECDM -62 08
\SIZE 17 \TYPE S
\FLUX1GHZ 5? \ALPHA 0.4?
\NAMES

\DATE 21 Aug 1996

\LONGITUDE 304.6 \LATITUDE +0.1
\RAHMS 13 05 59 \DECDM -62 42
\SIZE 8 \TYPE S
\FLUX1GHZ 14 \ALPHA 0.5
\NAMES Kes 17

\DATE 16 Jun 2017

\LONGITUDE 306.3 \LATITUDE -0.9
\RAHMS 13 21 50 \DECDM -63 34
\SIZE 4 \TYPE S?
\FLUX1GHZ 0.16? \ALPHA 0.5?
\NAMES

\DATE 9 Apr 2019

\LONGITUDE 308.1 \LATITUDE -0.7
\RAHMS 13 37 37 \DECDM -63 04
\SIZE 13 \TYPE S
\FLUX1GHZ 1.2? \ALPHA ?
\NAMES

\DATE 21 Aug 1996

\LONGITUDE 308.4 \LATITUDE -1.4
\RAHMS 13 41 30 \DECDM -63 44
\SIZE 12\X/6? \TYPE S?
\FLUX1GHZ 0.4? \ALPHA ?
\NAMES

\DATE 7 May 2014

\LONGITUDE 308.8 \LATITUDE -0.1
\RAHMS 13 42 30 \DECDM -62 23
\SIZE 30\X/20? \TYPE C?
\FLUX1GHZ 15? \ALPHA 0.4?
\NAMES

\DATE 5 Jun 2019

\LONGITUDE 309.2 \LATITUDE -0.6
\RAHMS 13 46 31 \DECDM -62 54
\SIZE 15\X/12 \TYPE S
\FLUX1GHZ 7? \ALPHA 0.4?
\NAMES

\DATE 27 Mar 2009

\LONGITUDE 309.8 \LATITUDE +0.0
\RAHMS 13 50 30 \DECDM -62 05
\SIZE 25\X/19 \TYPE S
\FLUX1GHZ 17 \ALPHA 0.5
\NAMES

\DATE 13 Aug 1998

\LONGITUDE 310.6 \LATITUDE -1.6
\RAHMS 14 00 45 \DECDM -63 26
\SIZE 2.5 \TYPE C?
\FLUX1GHZ ? \ALPHA ?
\NAMES

\DATE 4 Jun 2017

\LONGITUDE 310.6 \LATITUDE -0.3
\RAHMS 13 58 00 \DECDM -62 09
\SIZE 8 \TYPE S
\FLUX1GHZ 5? \ALPHA ?
\NAMES Kes 20B

\DATE 9 Apr 2019

\LONGITUDE 310.8 \LATITUDE -0.4
\RAHMS 14 00 00 \DECDM -62 17
\SIZE 12 \TYPE S
\FLUX1GHZ 6? \ALPHA ?
\NAMES Kes 20A

\DATE 9 Apr 2019

\LONGITUDE 311.5 \LATITUDE -0.3
\RAHMS 14 05 38 \DECDM -61 58
\SIZE 5 \TYPE S
\FLUX1GHZ 3? \ALPHA 0.5
\NAMES

\DATE 9 Apr 2019

\LONGITUDE 312.4 \LATITUDE -0.4
\RAHMS 14 13 00 \DECDM -61 44
\SIZE 38 \TYPE S
\FLUX1GHZ 45 \ALPHA 0.36
\NAMES

\DATE 9 Apr 2019

\LONGITUDE 312.5 \LATITUDE -3.0
\RAHMS 14 21 00 \DECDM -64 12
\SIZE 20\X/18 \TYPE S
\FLUX1GHZ 3.5? \ALPHA ?
\NAMES

\DATE 24 Apr 2014

\LONGITUDE 315.1 \LATITUDE +2.7
\RAHMS 14 24 30 \DECDM -57 50
\SIZE 190\X/150 \TYPE S
\FLUX1GHZ ? \ALPHA ?
\NAMES

\DATE 20 May 2014

\LONGITUDE 315.4 \LATITUDE -2.3
\RAHMS 14 43 00 \DECDM -62 30
\SIZE 42 \TYPE S
\FLUX1GHZ 49 \ALPHA 0.6
\NAMES RCW 86, MSH 14$-$6{\sl 3}

\DATE 15 Apr 2019

\LONGITUDE 315.4 \LATITUDE -0.3
\RAHMS 14 35 55 \DECDM -60 36
\SIZE 24\X/13 \TYPE ?
\FLUX1GHZ 8 \ALPHA 0.4
\NAMES

\DATE 25 Apr 2014

\LONGITUDE 315.9 \LATITUDE -0.0
\RAHMS 14 38 25 \DECDM -60 11
\SIZE 25\X/14 \TYPE S
\FLUX1GHZ 0.8? \ALPHA ?
\NAMES

\DATE 24 Apr 2014

\LONGITUDE 316.3 \LATITUDE -0.0
\RAHMS 14 41 30 \DECDM -60 00
\SIZE 29\X/14 \TYPE S
\FLUX1GHZ 20? \ALPHA 0.4
\NAMES (MSH 14$-$5{\sl 7})

\DATE 13 May 2017

\LONGITUDE 317.3 \LATITUDE -0.2
\RAHMS 14 49 40 \DECDM -59 46
\SIZE 11 \TYPE S
\FLUX1GHZ 4.7? \ALPHA ?
\NAMES

\DATE 25 Apr 2014

\LONGITUDE 318.2 \LATITUDE +0.1
\RAHMS 14 54 50 \DECDM -59 04
\SIZE 40\X/35 \TYPE S
\FLUX1GHZ $>$3.9? \ALPHA ?
\NAMES

\DATE 13 May 2017

\LONGITUDE 318.9 \LATITUDE +0.4
\RAHMS 14 58 30 \DECDM -58 29
\SIZE 30\X/14 \TYPE C
\FLUX1GHZ 4? \ALPHA 0.2?
\NAMES

\DATE 13 Aug 1998

\LONGITUDE 320.4 \LATITUDE -1.2
\RAHMS 15 14 30 \DECDM -59 08
\SIZE 35 \TYPE C
\FLUX1GHZ 60? \ALPHA 0.4
\NAMES MSH 15$-$5{\sl 2}, RCW 89

\DATE 15 Apr 2019

\LONGITUDE 320.6 \LATITUDE -1.6
\RAHMS 15 17 50 \DECDM -59 16
\SIZE 60\X/30 \TYPE S
\FLUX1GHZ ? \ALPHA ?
\NAMES

\DATE 25 Apr 2014

\LONGITUDE 321.9 \LATITUDE -1.1
\RAHMS 15 23 45 \DECDM -58 13
\SIZE 28 \TYPE S
\FLUX1GHZ $>$3.4? \ALPHA ?
\NAMES

\DATE 7 Jan 2004

\LONGITUDE 321.9 \LATITUDE -0.3
\RAHMS 15 20 40 \DECDM -57 34
\SIZE 31\X/23 \TYPE S
\FLUX1GHZ 13 \ALPHA 0.3
\NAMES

\DATE 8 Jun 2019

\LONGITUDE 322.1 \LATITUDE +0.0
\RAHMS 15 20 49 \DECDM -57 10
\SIZE 8\X/4.5? \TYPE S?
\FLUX1GHZ ? \ALPHA ?
\NAMES

\DATE 30 Apr 2014

\LONGITUDE 322.5 \LATITUDE -0.1
\RAHMS 15 23 23 \DECDM -57 06
\SIZE 15 \TYPE C
\FLUX1GHZ 1.5 \ALPHA 0.4
\NAMES

\DATE 13 Aug 1998

\LONGITUDE 323.5 \LATITUDE +0.1
\RAHMS 15 28 42 \DECDM -56 21
\SIZE 13 \TYPE S
\FLUX1GHZ 3? \ALPHA 0.4?
\NAMES

\DATE 16 Mar 2009

\LONGITUDE 323.7 \LATITUDE -1.0
\RAHMS 15 34 30 \DECDM -57 12
\SIZE 51\X/38 \TYPE S
\FLUX1GHZ ? \ALPHA ?
\NAMES

\DATE 16 Apr 2019

\LONGITUDE 326.3 \LATITUDE -1.8
\RAHMS 15 53 00 \DECDM -56 10
\SIZE 38 \TYPE C
\FLUX1GHZ 145 \ALPHA varies
\NAMES MSH 15$-$5{\sl 6}

\DATE 9 Apr 2019

\LONGITUDE 327.1 \LATITUDE -1.1
\RAHMS 15 54 25 \DECDM -55 09
\SIZE 18 \TYPE C
\FLUX1GHZ 7? \ALPHA ?
\NAMES

\DATE 23 Apr 2019

\LONGITUDE 327.2 \LATITUDE -0.1
\RAHMS 15 50 55 \DECDM -54 18
\SIZE 5 \TYPE S
\FLUX1GHZ 0.4 \ALPHA ?
\NAMES

\DATE 20 Apr 2019

\LONGITUDE 327.4 \LATITUDE +0.4
\RAHMS 15 48 20 \DECDM -53 49
\SIZE 21 \TYPE S
\FLUX1GHZ 30? \ALPHA 0.6
\NAMES Kes 27

\DATE 12 Jun 2017

\LONGITUDE 327.4 \LATITUDE +1.0
\RAHMS 15 46 48 \DECDM -53 20
\SIZE 14 \TYPE S
\FLUX1GHZ 1.9? \ALPHA ?
\NAMES

\DATE 10 Oct 2001

\LONGITUDE 327.6 \LATITUDE +14.6
\RAHMS 15 02 50 \DECDM -41 56
\SIZE 30 \TYPE S
\FLUX1GHZ 19 \ALPHA 0.6
\NAMES SN1006, PKS 1459$-$41

\DATE 9 Apr 2019

\LONGITUDE 328.4 \LATITUDE +0.2
\RAHMS 15 55 30 \DECDM -53 17
\SIZE 5 \TYPE F
\FLUX1GHZ 15 \ALPHA 0.0
\NAMES (MSH 15$-$5{\sl 7})

\DATE 27 Mar 2009

\LONGITUDE 329.7 \LATITUDE +0.4
\RAHMS 16 01 20 \DECDM -52 18
\SIZE 40\X/33 \TYPE S
\FLUX1GHZ $>$34? \ALPHA ?
\NAMES

\DATE 16 Mar 2009

\LONGITUDE 330.0 \LATITUDE +15.0
\RAHMS 15 10 00 \DECDM -40 00
\SIZE 180? \TYPE S
\FLUX1GHZ 350? \ALPHA 0.5?
\NAMES Lupus Loop

\DATE 30 May 2014

\LONGITUDE 330.2 \LATITUDE +1.0
\RAHMS 16 01 06 \DECDM -51 34
\SIZE 11 \TYPE S?
\FLUX1GHZ 5? \ALPHA 0.3
\NAMES

\DATE 9 Apr 2019

\LONGITUDE 332.0 \LATITUDE +0.2
\RAHMS 16 13 17 \DECDM -50 53
\SIZE 12 \TYPE S
\FLUX1GHZ 8? \ALPHA 0.5
\NAMES

\DATE 13 May 2017

\LONGITUDE 332.4 \LATITUDE -0.4
\RAHMS 16 17 33 \DECDM -51 02
\SIZE 10 \TYPE S
\FLUX1GHZ 28 \ALPHA 0.5
\NAMES RCW 103

\DATE 9 Apr 2019

\LONGITUDE 332.4 \LATITUDE +0.1
\RAHMS 16 15 20 \DECDM -50 42
\SIZE 15 \TYPE S
\FLUX1GHZ 26 \ALPHA 0.5
\NAMES MSH 16$-$5{\sl 1}, Kes 32

\DATE 30 May 2014

\LONGITUDE 332.5 \LATITUDE -5.6
\RAHMS 16 43 20 \DECDM -54 30
\SIZE 35 \TYPE S
\FLUX1GHZ 2? \ALPHA 0.7?
\NAMES

\DATE 9 Apr 2019

\LONGITUDE 335.2 \LATITUDE +0.1
\RAHMS 16 27 45 \DECDM -48 47
\SIZE 21 \TYPE S
\FLUX1GHZ 16 \ALPHA 0.5
\NAMES

\DATE 3 Apr 2019

\LONGITUDE 336.7 \LATITUDE +0.5
\RAHMS 16 32 11 \DECDM -47 19
\SIZE 14\X/10 \TYPE S
\FLUX1GHZ 6 \ALPHA 0.5
\NAMES

\DATE 25 Apr 2014

\LONGITUDE 337.0 \LATITUDE -0.1
\RAHMS 16 35 57 \DECDM -47 36
\SIZE 1.5 \TYPE S
\FLUX1GHZ 1.5 \ALPHA 0.6?
\NAMES (CTB 33)

\DATE 23 May 2014

\LONGITUDE 337.2 \LATITUDE -0.7
\RAHMS 16 39 28 \DECDM -47 51
\SIZE 6 \TYPE S
\FLUX1GHZ 1.5 \ALPHA 0.4
\NAMES

\DATE 13 May 2017

\LONGITUDE 337.2 \LATITUDE +0.1
\RAHMS 16 35 55 \DECDM -47 20
\SIZE 3\X/2 \TYPE ?
\FLUX1GHZ 1.5? \ALPHA ?
\NAMES

\DATE 19 Feb 2009

\LONGITUDE 337.3 \LATITUDE +1.0
\RAHMS 16 32 39 \DECDM -46 36
\SIZE 15\X/12 \TYPE S
\FLUX1GHZ 16 \ALPHA 0.55
\NAMES Kes 40

\DATE 13 Aug 1998

\LONGITUDE 337.8 \LATITUDE -0.1
\RAHMS 16 39 01 \DECDM -46 59
\SIZE 9\X/6 \TYPE S
\FLUX1GHZ 15 \ALPHA 0.5
\NAMES Kes 41

\DATE 9 Apr 2019

\LONGITUDE 338.1 \LATITUDE +0.4
\RAHMS 16 37 59 \DECDM -46 24
\SIZE 15? \TYPE S
\FLUX1GHZ 4? \ALPHA 0.4
\NAMES

\DATE 28 Apr 2017

\LONGITUDE 338.3 \LATITUDE -0.0
\RAHMS 16 41 00 \DECDM -46 34
\SIZE 8 \TYPE C?
\FLUX1GHZ 7? \ALPHA ?
\NAMES

\DATE 9 Apr 2019

\LONGITUDE 338.5 \LATITUDE +0.1
\RAHMS 16 41 09 \DECDM -46 19
\SIZE 9 \TYPE ?
\FLUX1GHZ 12? \ALPHA ?
\NAMES

\DATE 9 Apr 2019

\LONGITUDE 340.4 \LATITUDE +0.4
\RAHMS 16 46 31 \DECDM -44 39
\SIZE 10\X/7 \TYPE S
\FLUX1GHZ 5 \ALPHA 0.4
\NAMES

\DATE 25 Apr 2014

\LONGITUDE 340.6 \LATITUDE +0.3
\RAHMS 16 47 41 \DECDM -44 34
\SIZE 6 \TYPE S
\FLUX1GHZ 5? \ALPHA 0.4?
\NAMES

\DATE 22 Apr 2014

\LONGITUDE 341.2 \LATITUDE +0.9
\RAHMS 16 47 35 \DECDM -43 47
\SIZE 22\X/16 \TYPE C
\FLUX1GHZ 1.5? \ALPHA 0.6?
\NAMES

\DATE 8 Nov 2001

\LONGITUDE 341.9 \LATITUDE -0.3
\RAHMS 16 55 01 \DECDM -44 01
\SIZE 7 \TYPE S
\FLUX1GHZ 2.5 \ALPHA 0.5
\NAMES

\DATE 27 Jan 2004

\LONGITUDE 342.0 \LATITUDE -0.2
\RAHMS 16 54 50 \DECDM -43 53
\SIZE 12\X/9 \TYPE S
\FLUX1GHZ 3.5? \ALPHA 0.4?
\NAMES

\DATE 1 Aug 2000

\LONGITUDE 342.1 \LATITUDE +0.9
\RAHMS 16 50 43 \DECDM -43 04
\SIZE 10\X/9 \TYPE S
\FLUX1GHZ 0.5? \ALPHA ?
\NAMES

\DATE 13 Aug 1998

\LONGITUDE 343.0 \LATITUDE -6.0
\RAHMS 17 25 00 \DECDM -46 30
\SIZE 250 \TYPE S
\FLUX1GHZ ? \ALPHA ?
\NAMES RCW 114

\DATE 22 Apr 2014

\LONGITUDE 343.1 \LATITUDE -2.3
\RAHMS 17 08 00 \DECDM -44 16
\SIZE 32? \TYPE C?
\FLUX1GHZ 8? \ALPHA 0.5?
\NAMES

\DATE 22 Apr 2014

\LONGITUDE 343.1 \LATITUDE -0.7
\RAHMS 17 00 25 \DECDM -43 14
\SIZE 27\X/21 \TYPE S
\FLUX1GHZ 7.8 \ALPHA 0.55
\NAMES

\DATE 1 Aug 2000

\LONGITUDE 344.7 \LATITUDE -0.1
\RAHMS 17 03 51 \DECDM -41 42
\SIZE 8 \TYPE C?
\FLUX1GHZ 2.5? \ALPHA 0.3?
\NAMES

\DATE 23 May 2014

\LONGITUDE 345.7 \LATITUDE -0.2
\RAHMS 17 07 20 \DECDM -40 53
\SIZE 6 \TYPE S
\FLUX1GHZ 0.6? \ALPHA ?
\NAMES

\DATE 13 Aug 1998

\LONGITUDE 346.6 \LATITUDE -0.2
\RAHMS 17 10 19 \DECDM -40 11
\SIZE 8 \TYPE S
\FLUX1GHZ 8? \ALPHA 0.5?
\NAMES

\DATE 9 Apr 2019

\LONGITUDE 347.3 \LATITUDE -0.5
\RAHMS 17 13 50 \DECDM -39 45
\SIZE 65\X/55 \TYPE S?
\FLUX1GHZ 30? \ALPHA ?
\NAMES RX J1713.7$-$3946

\DATE 15 Apr 2019

\LONGITUDE 348.5 \LATITUDE -0.0
\RAHMS 17 15 26 \DECDM -38 28
\SIZE 10? \TYPE S?
\FLUX1GHZ 10? \ALPHA 0.4?
\NAMES

\DATE 5 Jun 2019

\LONGITUDE 348.5 \LATITUDE +0.1
\RAHMS 17 14 06 \DECDM -38 32
\SIZE 15 \TYPE S
\FLUX1GHZ 72 \ALPHA 0.3
\NAMES CTB 37A

\DATE 27 Jun 2017

\LONGITUDE 348.7 \LATITUDE +0.3
\RAHMS 17 13 55 \DECDM -38 11
\SIZE 17? \TYPE S
\FLUX1GHZ 26 \ALPHA 0.3
\NAMES CTB 37B

\DATE 13 May 2017

\LONGITUDE 349.2 \LATITUDE -0.1
\RAHMS 17 17 15 \DECDM -38 04
\SIZE 9\X/6 \TYPE S
\FLUX1GHZ 1.4? \ALPHA ?
\NAMES

\DATE 27 Aug 1996

\LONGITUDE 349.7 \LATITUDE +0.2
\RAHMS 17 17 59 \DECDM -37 26
\SIZE 2.5\X/2 \TYPE S
\FLUX1GHZ 20 \ALPHA 0.5
\NAMES

\DATE 15 Apr 2019

\LONGITUDE 350.0 \LATITUDE -2.0
\RAHMS 17 27 50 \DECDM -38 32
\SIZE 45 \TYPE S
\FLUX1GHZ 26 \ALPHA 0.4
\NAMES

\DATE 13 May 2017

\LONGITUDE 350.1 \LATITUDE -0.3
\RAHMS 17 21 05 \DECDM -37 27
\SIZE 4? \TYPE ?
\FLUX1GHZ 6? \ALPHA 0.8?
\NAMES

\DATE 5 May 2017

\LONGITUDE 351.0 \LATITUDE -5.4
\RAHMS 17 46 00 \DECDM -39 25
\SIZE 30 \TYPE S
\FLUX1GHZ ? \ALPHA ?
\NAMES

\DATE 12 Jun 2017

\LONGITUDE 351.2 \LATITUDE +0.1
\RAHMS 17 22 27 \DECDM -36 11
\SIZE 7 \TYPE C?
\FLUX1GHZ 5? \ALPHA 0.4
\NAMES

\DATE 13 Aug 1998

\LONGITUDE 351.7 \LATITUDE +0.8
\RAHMS 17 21 00 \DECDM -35 27
\SIZE 18\X/14 \TYPE S
\FLUX1GHZ 10 \ALPHA 0.5?
\NAMES

\DATE 19 Feb 2009

\LONGITUDE 351.9 \LATITUDE -0.9
\RAHMS 17 28 52 \DECDM -36 16
\SIZE 12\X/9 \TYPE S
\FLUX1GHZ 1.8? \ALPHA ?
\NAMES

\DATE 21 Aug 1996

\LONGITUDE 352.7 \LATITUDE -0.1
\RAHMS 17 27 40 \DECDM -35 07
\SIZE 8\X/6 \TYPE S
\FLUX1GHZ 4 \ALPHA 0.6
\NAMES

\DATE 5 May 2017

\LONGITUDE 353.6 \LATITUDE -0.7
\RAHMS 17 32 00 \DECDM -34 44
\SIZE 30 \TYPE S
\FLUX1GHZ 2.5? \ALPHA ?
\NAMES

\DATE 15 Apr 2019

\LONGITUDE 353.9 \LATITUDE -2.0
\RAHMS 17 38 55 \DECDM -35 11
\SIZE 13 \TYPE S
\FLUX1GHZ 1? \ALPHA 0.5?
\NAMES

\DATE 8 Nov 2001

\LONGITUDE 354.1 \LATITUDE +0.1
\RAHMS 17 30 28 \DECDM -33 46
\SIZE 15\X/3? \TYPE C?
\FLUX1GHZ ? \ALPHA varies
\NAMES

\DATE 4 Jun 2017

\LONGITUDE 354.8 \LATITUDE -0.8
\RAHMS 17 36 00 \DECDM -33 42
\SIZE 19 \TYPE S
\FLUX1GHZ 2.8? \ALPHA ?
\NAMES

\DATE 1 Aug 2000

\LONGITUDE 355.4 \LATITUDE +0.7
\RAHMS 17 31 20 \DECDM -32 26
\SIZE 25 \TYPE S
\FLUX1GHZ 5? \ALPHA ?
\NAMES

\DATE 4 Jun 2017

\LONGITUDE 355.6 \LATITUDE -0.0
\RAHMS 17 35 16 \DECDM -32 38
\SIZE 8\X/6 \TYPE S
\FLUX1GHZ 3? \ALPHA ?
\NAMES

\DATE 25 Apr 2014

\LONGITUDE 355.9 \LATITUDE -2.5
\RAHMS 17 45 53 \DECDM -33 43
\SIZE 13 \TYPE S
\FLUX1GHZ 8 \ALPHA 0.5
\NAMES

\DATE 25 Apr 2014

\LONGITUDE 356.2 \LATITUDE +4.5
\RAHMS 17 19 00 \DECDM -29 40
\SIZE 25 \TYPE S
\FLUX1GHZ 4 \ALPHA 0.7
\NAMES

\DATE 27 Apr 2014

\LONGITUDE 356.3 \LATITUDE -1.5
\RAHMS 17 42 35 \DECDM -32 52
\SIZE 20\X/15 \TYPE S
\FLUX1GHZ 3? \ALPHA ?
\NAMES

\DATE 20 May 2014

\LONGITUDE 356.3 \LATITUDE -0.3
\RAHMS 17 37 56 \DECDM -32 16
\SIZE 11\X/7 \TYPE S
\FLUX1GHZ 3? \ALPHA ?
\NAMES

\DATE 13 May 2017

\LONGITUDE 357.7 \LATITUDE -0.1
\RAHMS 17 40 29 \DECDM -30 58
\SIZE 8\X/3? \TYPE ?
\FLUX1GHZ 37 \ALPHA 0.4
\NAMES MSH 17$-$3{\sl 9}

\DATE 9 Apr 2019

\LONGITUDE 357.7 \LATITUDE +0.3
\RAHMS 17 38 35 \DECDM -30 44
\SIZE 24 \TYPE S
\FLUX1GHZ 10 \ALPHA 0.4?
\NAMES

\DATE 9 Apr 2019

\LONGITUDE 358.0 \LATITUDE +3.8
\RAHMS 17 26 00 \DECDM -28 36
\SIZE 38 \TYPE S
\FLUX1GHZ 1.5? \ALPHA ?
\NAMES

\DATE 27 Apr 2014

\LONGITUDE 358.1 \LATITUDE +1.0
\RAHMS 17 37 00 \DECDM -29 59
\SIZE 20 \TYPE S
\FLUX1GHZ 2? \ALPHA ?
\NAMES

\DATE 28 Apr 2017

\LONGITUDE 358.5 \LATITUDE -0.9
\RAHMS 17 46 10 \DECDM -30 40
\SIZE 17 \TYPE S
\FLUX1GHZ 4? \ALPHA ?
\NAMES

\DATE 16 Mar 2009

\LONGITUDE 359.0 \LATITUDE -0.9
\RAHMS 17 46 50 \DECDM -30 16
\SIZE 23 \TYPE S
\FLUX1GHZ 23 \ALPHA 0.5
\NAMES

\DATE 3 Apr 2019

\LONGITUDE 359.1 \LATITUDE -0.5
\RAHMS 17 45 30 \DECDM -29 57
\SIZE 24 \TYPE S
\FLUX1GHZ 14 \ALPHA 0.4?
\NAMES

\DATE 9 Apr 2019

\LONGITUDE 359.1 \LATITUDE +0.9
\RAHMS 17 39 36 \DECDM -29 11
\SIZE 12\X/11 \TYPE S
\FLUX1GHZ 2? \ALPHA ?
\NAMES

\DATE 19 Feb 2009

%% file: arxiv2.bbl
\begin{thebibliography}{10}

\bibitem[Ackermann et al.(2018)Ackermann et al.]{2018ApJS..237...32A}
  Ackermann M., et al.,
    2018, ApJS, 237, 32

\bibitem[Anderson et al.(2017)Anderson et al.]{2017A&A...605A..58A}
  Anderson L.~D., et al.,
    2017, A\&A, 605, A58

\bibitem[Araya(2017)Araya]{2017ApJ...843...12A}
  Araya M.,
    2017, ApJ, 843, 12

\bibitem[Beuther et al.(2016)Beuther et al.]{2016A&A...595A..32B}
  Beuther H., et al.,
    2016, A\&A, 595, A32

\bibitem[Boumis et al.(2002)Boumis et al.]{2002A&A...396..225B}
  Boumis P., Mavromatakis F., Paleologou E.~V., Becker W.,
    2002, A\&A, 396, 225

\bibitem[Condon et al.(1998)Condon et al.]{1998AJ....115.1693C}
  Condon J.~J., Cotton W.~D., Greisen E.~W., Yin Q.~F., Perley R.~A.,
  Taylor G.~B., Broderick J.~J.,
    1998, AJ, 115, 1693

\bibitem[de Gasperin et al.(2014)de Gasperin et al.]{2014A&A...568A.107D}
  de Gasperin F., Evoli C., Br{\"u}ggen M., Hektor A., Cardillo M.,
  Thorman P., Dawson W.~A., Morrison C.~B.,
    2014, A\&A, 568, A107

\bibitem[Demetroullas et al.(2015)Demetroullas et al.]{2015MNRAS.453.2082D}
  Demetroullas C., et al.,
    2015, MNRAS, 453, 2082

\bibitem[Dokara et al.(2018)Dokara et al.]{2018ApJ...866...61D}
  Dokara, R., et al.,
    2018, ApJ, 866, 61

\bibitem[Driessen et al.(2018)Driessen et al.]{2018ApJ...860..133D}
  Driessen L.~N., Dom{\v c}ek V., Vink J., Hessels J.~W.~T., Arias M., Gelfand J.~D.,
    2018, ApJ, 860, 133

\bibitem[Dzib et al.(2018)Dzib et al.]{2018ApJ...866..100D}
  Dzib S.~A., Rodr{\'{\i}}guez L.~F., Karuppusamy R., Loinard L., Medina S.-N.~X.,
    2018, ApJ, 866, 100

\bibitem[Ferrand \& Safi-Harb(2012)Ferrand \& Safi-Harb]{2012AdSpR..49.1313F}
  Ferrand, G., \& Safi-Harb, S.,
    2012, AdSpR, 49, 1313

\bibitem[Fesen et al.(2015)Fesen et al.]{2015ApJ...812...37F}
  Fesen R.~A., Neustadt J.~M.~M., Black C.~S., Koeppel A.~H.~D.,
    2015, ApJ, 812, 37

\bibitem[Froebrich et al.(2015)Froebrich et al.]{2015MNRAS.454.2586F}
  Froebrich D., et al.,
    2015, MNRAS, 454, 2586

\bibitem[Gao \& Han(2014)Gao \& Han]{2014A&A...567A..59G}
  Gao X.~Y., Han J.~L.,
    2014, A\&A, 567, A59

\bibitem[Gao et al.(2011)Gao et al.]{2011A&A...529A.159G}
  Gao X.~Y., Han J.~L., Reich W., Reich P., Sun X.~H., Xiao L.,
    2011, A\&A, 529, A159

\bibitem[Gerbrandt et al.(2014)Gerbrandt et al.]{2014A&A...566A..76G}
  Gerbrandt S., Foster T.~J., Kothes R., Geisb{\"u}sch J., Tung A.,
    2014, A\&A, 566, A76

\bibitem[Giveon et al.(2005)Giveon et al.]{2005AJ....129..348G}
  Giveon U., Becker R.~H., Helfand D.~J., White R.~L.,
    2005, AJ, 129, 348

\bibitem[Green(1984)Green]{1984MNRAS.209..449G}
  Green D.~A.,
    1984, MNRAS, 209, 449

\bibitem[Green(1988)Green]{1988Ap&SS.148....3G}
  Green D.~A.,
    1988, Ap\&SS, 148, 3

\bibitem[Green(1990)Green]{1990AJ....100.1927G}
  Green D.~A.,
    1990, AJ, 100, 192

\bibitem[Green(1991)Green]{1991PASP..103..209G}
  Green D.~A.,
    1991, PASP, 103, 209

\bibitem[Green(2004)Green]{2004BASI...32..335G}
  Green D.~A.,
    2004, BASI, 32, 335

\bibitem[Green(2005)Green]{2005MmSAI..76..534G}
  Green D.~A.,
    2005, MmSAI, 76, 534

\bibitem[Green(2009)Green]{2009BASI...37...45G}
  Green D.~A.,
    2009, BASI, 37, 45

\bibitem[Green(2014)Green]{2014BASI...42...47G}
  Green D.~A.,
    2014, BASI, 42, 47

\bibitem[Green(2015)Green]{2015MNRAS.454.1517G}
  Green D.~A.,
    2015, MNRAS, 454, 1517

\bibitem[Green et al.(2008)Green et al.]{2008MNRAS.387L..54G}
  Green D.~A., Reynolds S.~P., Borkowski K.~J., Hwang U., Harrus I., Petre R.,
    2008, MNRAS, 387, L54

\bibitem[Green et al.(2014)Green, Reeves \& Murphy]{2014PASA...31...42G}
  Green A.~J., Reeves S.~N., Murphy T.,
    2014, PASA, 31, e042

\bibitem[Haverkorn et al.(2006)Haverkorn et al.]{2006ApJS..167..230H}
  Haverkorn M., Gaensler B.~M., McClure-Griffiths N.~M., Dickey J.~M., Green A.~J.,
    2006, ApJS, 167, 230

\bibitem[H.E.S.S.~Collaboration et al.(2018)H.E.S.S.~Collaboration et al.]{2018A&A...612A...8H}
  H.E.S.S.~Collaboration (Abdalla H., et al.),
    2018, A\&A, 612, A8

\bibitem[Helfand et al.(2006)Helfand et al.]{2006AJ....131.2525H}
  Helfand D.~J., Becker R.~H., White R.~L., Fallon A., Tuttle S.,
    2006, AJ, 131, 2525

\bibitem[Kang et al.(2014)Kang, Koo, \& Byun]{2014JKAS...47..259K}
  Kang J.-H., Koo B.-C., Byun D.-Y.,
    2014, JKAS, 47, 259

\bibitem[Kothes et al.(2017)Kothes et al.]{2017A&A...597A.116K}
  Kothes R., Reich P., Foster T.~J., Reich W.,
    2017, A\&A, 597, A116

\bibitem[Murphy et al.(2007)Murphy et al.]{2007MNRAS.382..382M}
  Murphy T., Mauch T., Green A., Hunstead R.~W., Piestrzynska B., Kels A.~P., Sztajer P.,
    2007, MNRAS, 382, 382

\bibitem[Perley \& Butler(2017)Perley \& Butler]{2017ApJS..230....7P}
  Perley R.~A., Butler B.~J.,
    2017, ApJS, 230, 7

\bibitem[Proctor(2016)Proctor]{2016ApJS..224...18P}
  Proctor D.~D.,
    2016, ApJS, 224, 18

\bibitem[Reynolds et al.(2008)Reynolds et al.]{2008ApJ...680L..41R}
  Reynolds S.~P., Borkowski K.~J., Green D.~A., Hwang U., Harrus I., Petre R.,
    2008, ApJ, 680, L41

\bibitem[Rho \& Petre(1998)Rho \& Petre]{1998ApJ...503L.167R}
  Rho J., Petre R.,
    1998, ApJ, 503, L167

\bibitem[Stephenson \& Green(2002)Stephenson \& Green]{2002ISAA....5.....S}
  Stephenson F.~R., Green D.~A., 2002, Historical supernovae and their
  remnants, Oxford University Press

\bibitem[Stil et al.(2006)Stil et al.]{2006AJ....132.1158S}
  Stil J.~M., et al.,
    2006, AJ, 132, 1158

\bibitem[Xu et al.(2013)Xu et al.]{2013A&A...559A..81X}
  Xu W.~F., Gao X.~Y., Han J.~L., Liu F.~S.,
    2013, A\&A, 559, A81

\end{thebibliography}
